\documentclass[a4paper,12pt]{iopart}
\pdfoutput=1
\usepackage{textcomp}
\usepackage{amsmath}
\usepackage{amssymb}
\usepackage{graphicx} 
\usepackage{pgf}
\def\Xint#1{\mathchoice
{\XXint\displaystyle\textstyle{#1}} 
{\XXint\textstyle\scriptstyle{#1}} 
{\XXint\scriptstyle\scriptscriptstyle{#1}} 
{\XXint\scriptscriptstyle\scriptscriptstyle{#1}} 
\!\int}
\def\XXint#1#2#3{{\setbox0=\hbox{$#1{#2#3}{\int}$ }
\vcenter{\hbox{$#2#3$ }}\kern-.6\wd0}}

\def\dashint{\Xint-}

\newcommand{\ket}[1]{\left|#1\right\rangle}
\newcommand{\bra}[1]{\left\langle #1\right|}
\newcommand{\ud}{\mathrm{d}}

\newcommand{\mean}[1]{\left\langle #1\right\rangle}

\newcommand{\nep}{\textrm{e}}
\newcommand{\atan}{\operatorname{atan}}
\newcommand{\Real}{\Re e}

\newcommand{\bgamma}{\mbox{\boldmath $\gamma$}}
\begin{document}
 
\title[Linear response as a singular limit]{Linear response as a singular limit for a periodically driven closed quantum system}

\author{Angelo Russomanno$^{1,2}$, Alessandro Silva$^{1,3}$, and Giuseppe E. Santoro$^{1,2,3}$}

\address{$^1$ SISSA, Via Bonomea 265, I-34136 Trieste, Italy}
\address{$^2$ CNR-IOM Democritos National Simulation Center, Via Bonomea 265, I-34136 Trieste, Italy}
\address{$^3$ International Centre for Theoretical Physics (ICTP), P.O.Box 586, I-34014 Trieste, Italy}
\eads{\mailto{russoman@sissa.it}, \mailto{asilva@sissa.it}, \mailto{santoro@sissa.it}}

\begin{abstract}
We address the issue of the validity of linear response theory for a closed quantum system subject to a periodic external driving.
Linear response theory (LRT) predicts energy absorption at frequencies of the external driving where the imaginary part of the appropriate response function is different from zero. 
Here we show that, for a fairly general non-linear many-body system on a lattice subject to an extensive perturbation, this approximation should be expected to be valid 
only up to a time $t^*$ depending on the strength of the driving, beyond which the true coherent Schr\"odinger evolution departs 
from the linear response prediction and the system stops absorbing energy form the driving. 
We exemplify this phenomenon in detail with the example of a quantum Ising chain subject to a time-periodic modulation 
of the transverse field, by comparing an exact Floquet analysis with the standard results of LRT. 
In this context, we also show that if the perturbation is just local, the system is expected in the thermodynamic limit to keep absorbing energy, and LRT works at all times. 
We finally argue more generally the validity of the scenario presented for closed quantum many-body 
lattice systems with a bound on the energy-per-site spectrum, discussing the experimental relevance of our findings in the context of cold atoms in optical lattices and 
ultra-fast spectroscopy experiments.
\end{abstract}

\pacs{75.10.Pq, 05.30.Rt, 03.65.-w}

\maketitle
\tableofcontents 

\section{Introduction}

Linear response theory (LRT) is one most useful tools of statistical physics and condensed matter theory, both classical and quantum, 
treated in detail in most textbook \cite{Pines-Nozieres:book,Forster:book,Giuliani-Vignale:book}.
The success of Kubo formulas \cite{Kubo_JPSJ57} in describing the response of a system weakly perturbed out of equilibrium is 
well known. Its realm of application goes from transport coefficients in electronic systems \cite{Pines-Nozieres:book,Giuliani-Vignale:book} 
to relaxation phenomena in normal liquids, superfluids and magnetic system \cite{Forster:book}.

The theory, which is most easily formulated in the quantum case, expresses the response of the average value at time $t$ 
of an observable $\mean{{B}}_t$  for a system whose Hamiltonian ${H}$ is weakly perturbed by a term $v(t) {A}$ in terms of 
(retarded) response functions $\chi_{BA}$;  the $\chi_{BA}$'s, also known as susceptibilities, are in turn expressed 
in terms of equilibrium averages of commutators of the Heisenberg's operators ${B}_H(t)$ and ${A}_H(t')$,
where the time evolution is assumed to be perfectly unitary (coherent) and governed by the equilibrium Hamiltonian ${H}$.

One of the well known properties of LRT is that it predicts a response which is in general ``out-of-phase'' with the perturbation
--- the Fourier transformed susceptibilities $\chi(\omega)$ have imaginary parts --- and this is generally associated to {\em energy absorption}:
the systems takes energy from the driving forces at a positive rate controlled by the imaginary part of the appropriate response 
functions \cite{Pines-Nozieres:book,Forster:book,Giuliani-Vignale:book}.
 
Admittedly, some of the ingredients in the standard derivations of Kubo formulas --- like for instance the assumption, in the quantum case, of a
perfectly ``coherent'' evolution --- are not easy to justify, at least on the macroscopic time-scales over which the results are succesfully applied.
Van Kampen has even harshly criticized the whole theory as a ``mathematical exercise'' \cite{vanKampen_PN71} trying to bridge the
huge time-gap between the expected linearity at the macroscopic scale with an unjustified assumption of linearity in the microscopic equations of
motion. 
Even without taking such an extreme view --- after all, this ``mathematical exercise'' is remarkably successfull ---  one could still try to 
test the regime of validity of linear response in the time-domain, in a setting in which a coherent evolution is guaranteed:
this might apply both to experiments on cold atoms in optical lattices \cite{Bloch_RMP08} as well as to more conventional
condensed matter system studied by ultra-fast spectroscopies~\cite{Enciclopedia:book,Shah:book,Glezer:phdthesis,Nasu:book} 
where the dynamics of a system in the sub-pico-second range is likely not affected by the interaction with the environment.

An ideal testing ground for LRT is the coherent unitary evolution of a closed many-body quantum system subject to a periodic driving, 
where a Floquet analysis \cite{Shirley_PR65,Grifoni_PR98} can be applied provided the usual adiabatic switching-on factors 
are avoided. 
In a recent work ~\cite{Russomanno_PRL12}, we have considered such a problem for a one-dimensional
Ising model in a time-dependent uniform transverse field $h(t)$, and found that the response of the system to a periodic driving of
$h(t)$ results --- after a transient and in the thermodynamic limit --- in a periodic behaviour of the  
averages of the observables. When considering the transverse magnetization after the transient, in particular, this periodic behaviour
turned out to be  ``synchronized'' in-phase with the perturbating transverse field, in such a way as to have zero energy absorbed 
from the driving over a cycle. 
Though we have exemplified these ideas using a quantum Ising chain, we have argued for their more general validity under 
circumstances which could be fairly applicable to closed quantum many-body systems on a lattice
in absence of disorder \cite{Russomanno_PRL12}.
A question is however in order at this point: "synchronization" implies that the out-of-phase response typically associated (within LRT) 
to energy absorption and the imaginary part of the response functions, vanish. 
What is the physics behind this effect?
%
This is precisely the issue addressed by the present paper, where we plan to compare 
--- again, for definiteness, in the quantum Ising chain --- the results of an exact Floquet 
analysis in a regime of weak periodic driving of the transverse field with the outcome of LRT.
We consider both the case of a perturbation which is extensive, i.e., involving a number of sites $l$ which 
increases as the system size $L$ in the thermodynamic limit ($l, L \to \infty$ but $l/L\to \mbox{constant}$), 
as well as that of a local perturbation, where $l$ of order $1$.
For the case of an extensive perturbation, we find that the results of LRT are applicable only at short times, and
emerge from a rather singular limit in the strength of the perturbation. LRT would predict a
constant energy absorbtion at a rate proportional to the imaginary part of the corresponding response function. 
For any small but finite perturbation, the true response shows in turn the linear-in-time energy absorption predicted
by LRT only at short times, while eventually at longer times the true energy absorption rate vanishes. 
Correspondingly, of the two components of the LRT, ``in-phase'' and ``out-of-phase'' only the former survives in the asymptotic limit,
corresponding to the ``synchronization'' of the system with the perturbation. Interestingly, contrary to
the dissipative component, the strength of the ``in-phase'' response turns out to be well described by LRT.
%
%
This is essentially consistent with what is known in the context of mesoscopic physics about the origin of resistance and energy dissipation 
in small metallic loops subject to a time-dependent magnetic flux (a uniform electric field):
as discussed by Landauer \cite{Landauer_PRB86} and by Gefen and Thouless \cite{Gefen_PRL87}, due to phase coherence, 
Zener tunneling between bands does not imply energy dissipation but rather energy storage \cite{Landauer_PRB86},
and elastic scattering due to localized potentials will generally lead to a saturation of the energy absorbed by the system \cite{Gefen_PRL87}, 
without resistance (inelastic effects are essential for that).
In our case, we find that in order to describe accurately also the dissipative response with linear response theory 
the system should act ``as its own bath''. This happens, for example, when the weak perturbation/driving acts {\em locally} in a finite region $l$ of order $1$: 
in this case we find that LRT is essentially exact at all times $t$, as $L\to \infty$: the system can accomodate a linear-in-time energy increase (of order $1$) 
even for $t\to \infty$, as this adds a vanishingly small contribution to the energy-per-site, of order $1/L\to 0$.  

The rest of the paper is organized as follows. 
In Section \ref{linear-response:sec} we briefly review the LRT, for the reader's convenience, and present the slightly less 
common LRT calculation for a perfectly periodic perturbation without adiabatic switching-on factors. 
In Section \ref{floquet:sec} we present the Floquet analysis of a finite amplitude perturbation, and the arguments leading to an 
asymptotically periodic behaviour introduced in Ref.~\cite{Russomanno_PRL12}.
Section \ref{energy:sec} contains some general energy considerations leading to the conclusions that the true response 
often lacks the ``out-of-phase'' part predicted by LRT, in particular at least when the perturbation is extensive and the model has 
a finite bandwidth single-particle spectrum. 
In Section \ref{Ising:sec} we exemplify these general considerations with a quantum Ising chain subject to a time-periodic transverse field. 
We will show, Section \ref{comparison_l=L:sec}, that, while the LRT response proportional to the real part of $\chi(\omega)$ is perfectly matching the exact Floquet 
results for small driving, the out-of-phase response due to the imaginary part of $\chi(\omega)$ is, strictly speaking, missing 
at large times.
In Section \ref{subchain:sec} we discuss the case of a perturbation which extends spatially over a segment of the chain of length $l$, 
analysing the case in which $l/L$ remains constant in the thermodynamic limit (an extensive perturbation), contrasting it
with the case in which $l$ remains constant (a local perturbation), where LRT is asymptotically exact.
Section \ref{conclusions:sec} contains a summary of our results, a discussion of their experimental relevance, 
both for cold atoms in optical lattices and for ultra-fast spectroscopies, and our conclusions.
Four appendices contain some technical material on the analysis of the singularities of LRT, on the 
transverse magnetic susceptibility of the quantum Ising chain, and on the Bogoliubov-de Gennes-Floquet 
dynamics of a general inhomogeneous quantum Ising chain.

\section{Linear response theory} \label{linear-response:sec}
%
Let us start with a brief recap of LRT, as discussed in most textbooks~\cite{Pines-Nozieres:book,Forster:book,Giuliani-Vignale:book}.
Assume that the equilibrium Hamiltonian $\hat{H}_0$ of a given system is weakly perturbed
\begin{equation} \label{hammy}
  \hat{H}(t) = \hat{H}_0 + v(t)\hat{A} \;,
\end{equation}
where $\hat{A}$ is some Hermitean operator and $v(t)$ a (weak) perturbing field. 
At equilibrium, the system would be governed by a thermal (Gibbs) density matrix at 
a (possibly vanishing) temperature $T=1/(k_B\beta)$:
\begin{equation}
  \hat{\rho}_{\rm eq} = \sum_n \frac{ \nep^{-\beta E_n^{(0)}}}{Z} \ket{\Phi^{(0)}_n}\bra{\Phi^{(0)}_n} \;,
\end{equation}
where $\ket{\Phi^{(0)}_n}$ are the eigenstates of $\hat{H}_0$, $E_n^{(0)}$ the corresponding eigenenergies, 
and $Z= \sum_n \nep^{-\beta E_n^{(0)}}$ the partition sum.
LRT tells us how to calculate the (perturbed) expectation value of any operator $\hat{B}$ at time $t$, 
$\mean{B}_t$, to linear order in the perturbation $v(t)$, assuming a coherent (unitary) evolution governed by $\hat{H}(t)$. 
Restricting our considerations to the case $\hat{B}=\hat{A}$, we know that \cite{Giuliani-Vignale:book}: 
\begin{equation} \label{valtim}
  \mean{A}_t = \mean{A}_{\rm eq} + \int_{-\infty}^{+\infty}\!\! \ud t' \; \chi(t-t') \, v(t') \;,
\end{equation}
where $\mean{A}_{\rm eq}={\rm Tr}[\hat{\rho}_{\rm eq} \, \hat{A}]$ is the equilibrium value, and the retarded susceptibility $\chi(t)$ is given by:
\begin{equation} \label{chicchiricchi}
  \chi(t) \equiv -\frac{i}{\hbar} \theta(t) \mean{\left[\hat{A}(t),\hat{A}\right]}_{\rm eq}
  =-\frac{i}{\hbar} \theta(t) \sum_{n,m}\left(\rho_m-\rho_n\right) \left| A_{m n}\right|^2 \nep^{-i\omega_{n m}t} \;,
\end{equation}
with $\rho_n= \nep^{-\beta E_n^{(0)}}/Z$, $A_{m n}=\bra{\Phi^{(0)}_m} \hat{A} \ket{\Phi^{(0)}_n}$, and $\hbar \omega_{n m}=E_n^{(0)}-E_m^{(0)}$.
The relevant information on the susceptibility is contained in its spectral function 
\begin{equation} \label{chisec}
  \chi''(\omega) =-\frac{\pi}{\hbar} \sum_{n,m} \left(\rho_m-\rho_n\right) \left| A_{mn} \right|^2 \delta(\omega-\omega_{nm}) \;,
\end{equation}
which is, essentially, the imaginary part of the Fourier-transform $\chi(z)$ for $z=\omega+i\eta$, with $\eta\to 0^+$, 
\begin{equation} \label{chiz:eqn}
\chi(z) =  \int_{-\infty}^{+\infty}\!\! \ud t \; \chi(t) \; e^{iz t} = 
\int_{-\infty}^{+\infty} \; \frac{\ud\omega}{\pi} \frac{\chi''(\omega)}{\omega-z} \;,
\end{equation}
and is manifestly {\em odd}:  $\chi''(-\omega)=-\chi''(\omega)$. 
We will always assume (unless otherwise stated) that we are dealing with an extended system in the thermodynamic limit, 
so that $\chi''(\omega)$ is a smooth function of $\omega$, rather than a sum of discrete Dirac's delta functions.


Consider now the case of a perfectly periodic perturbation of frequency $\omega_0$, for definiteness $v_0 \sin{(\omega_0 t)}$.
The standard textbook approach would include an adiabatic switching-on of the perturbation from $-\infty$ to $0$, writing 
$v(t) =  v_0 \sin(\omega_0 t) \left[ \nep^{\eta t} \theta(-t) + \theta(t) \right]$, with a small positive $\eta$ which is eventually sent to $0$
at the end of the calculation. 
Since we are interested in comparing LRT with a Floquet approach, we insist on a strictly periodic perturbation turned-on at $t=0$, 
and take $v(t)=v_{\rm per}(t) = v_0  \theta(t) \sin{(\omega_0 t)}$.
The calculation of the response to such a perturbation is an elementary application of Eqs.~(\ref{valtim}-\ref{chisec}), and gives:
\begin{equation}  \label{meaner}
  \delta\mean{A}_t^{\rm per} = v_0 \int_{-\infty}^{+\infty} \! \frac{\ud\omega}{2\pi i} \; \; 
  \chi''(\omega) \left( \frac{\nep^{i\omega_0 t}-\nep^{-i\omega t}}{\omega + \omega_0}-\frac{\nep^{-i\omega_0 t}-\nep^{-i\omega t}}{\omega-\omega_0}\right) \;,
\end{equation}
where $\delta\mean{A}_t^{\rm per}=\mean{A}_t^{\rm per}- \mean{A}_{\rm eq}$.
(A word of caution on the notation: $\delta\mean{A}_t^{\rm per}$ is the response to the periodic driving $v_{\rm per}(t)$, but is not itself periodic in time.)
We notice that, although the usual $\pm i \eta$ factors do not appear anywhere, the integrand in Eq. \eqref{meaner} is regular, nothwithstanding 
the singular denominators at $\omega = \pm \omega_0$, because the limits for $\omega \to \pm \omega_0$ are finite: there is no need, therefore, 
for a Cauchy principal value prescription. 
If we split the two contributions appearing in the numerators, with $\nep^{\pm i\omega_0 t}$ and $\nep^{-i\omega t}$, into two separate integrals, however, 
the singularity of the two denominators at $\omega = \pm \omega_0$ will require a principal value prescription for both.
By using the fact that $\chi''(\omega)$ is odd, one readily finds:
\begin{equation} \label{meaner1}
  \delta\mean{A}_t^{\rm per} = v_0  \chi'(\omega_0) \sin{(\omega_0 t)} 
- 2 v_0 \omega_0 \dashint_{0}^{+\infty} \! \! \frac{\ud\omega}{\pi} \; \frac{\chi''(\omega)}{\omega^2-\omega_0^2} \sin{(\omega t)} \;,
\end{equation}
where we have introduced the Kramers-Kr\"onig transform $\chi'(\omega_0)$ 
\begin{equation} \label{chi-prime:eqn}
  \chi'(\omega_0) \equiv \dashint_{-\infty}^{+\infty} \frac{\ud\omega}{\pi} \frac{\chi''(\omega)}{\omega-\omega_0} \;,
\end{equation}
i.e., see Eq.~\eqref{chiz:eqn}, the real part of $\chi(\omega+i\eta)$ on the upper real axis \cite{Giuliani-Vignale:book}.

A few comments are in order here. 
The Riemann-Lebesgue lemma \cite{Bochner:book} states that Fourier transforms of a regular function  
${\tilde F}(\omega)$ (such that $| {\tilde F}(\omega) |$ be Lebesgue-integrable) approach $0$ for large times
\begin{equation} \label{RL:eqn}
F(t) = \int \!\! d\omega \; {\tilde F}(\omega) \; e^{-i\omega t}  \; \; \stackrel{\scriptscriptstyle t\to \infty} {\longrightarrow} \; \; 0  .
\end{equation}
%
Physically, this result follows from dephasing associated to the overlap of the rapidly oscillating (for large $t$) phase-factors  $e^{-i\omega t}$ 
weighting the ``smooth'' ${\tilde F}(\omega)$.
%
%
%
The frequency integral appearing in the second term of Eq.~\eqref{meaner1}, however, 
has a singularity at $\omega=\omega_0$, which should be treated by the principal value prescription whenever $\chi''(\omega_0)\neq 0$.
This singularity does not allow a straightforward application of the Riemann-Lebesgue lemma, and leads to a large-$t$ value of the integral which does not decay
to $0$. Indeed, as explained in \ref{singularity-LR:app}, it is a simple matter to ``extract'' the singularity from the integral, by
isolating a term proportional to $\chi''(\omega_0)$,
which turns out to have the familiar form $-v_0 \chi''(\omega_0) \cos{(\omega_0 t)}$, plus a regular (transient) term $F^{\rm trans}(t)$, 
ending up with the expression:
\begin{equation} \label{LR-transient:eqn}
   \delta\mean{A}_t^{\rm per} = v_0 [  \chi'(\omega_0) \sin{(\omega_0 t)}  - \chi''(\omega_0) \cos{(\omega_0 t)}   ] + F^{\rm trans}(\omega_0,t)  \;,
\end{equation}
where the transient part 
\begin{equation}  \label{LR-transient-2:eqn}
F^{\rm trans}(\omega_0,t) = 
  -v_0 \int_{-\infty}^\infty \! \! \frac{\ud\omega}{\pi} \; \frac{[\chi''(\omega)-\chi''(\omega_0)]}{\omega-\omega_0} \sin{(\omega t)}  \;,
\end{equation}
is now vanishing for large $t$, due to the Riemann-Lebesgue lemma.
Therefore, LRT predicts a periodic response composed, at large times of two terms: one in-phase with the perturbation, proportional
to $\chi'(\omega_0)$, and one out-of-phase with it, proportional to $\chi''(\omega_0)$, associated to energy absorption (see below Section~\ref{energy:sec}). 

\section{Floquet theory and synchronization} \label{floquet:sec}
%
Let us now discuss the case of a periodic perturbation with a finite, but not necessarily small, amplitude. 
As in~\cite{Russomanno_PRL12}, the dynamics in this case case can be studied using Floquet theory. 
Let us now, to set the notation, briefly review the basics of Floquet theory, referring the reader to the available 
literature for more details~\cite{Shirley_PR65, Grifoni_PR98, Russomanno_PRL12,Russomanno_PRB11}. 
In case of a time periodic Hamiltonian like Eq.~\eqref{hammy}, i.e., $\hat{H}\left(t\right)=\hat{H}\left(t+\tau\right)$ with $\tau=2\pi/\omega_0$, 
in analogy with Bloch theorem in the standard band theory of crystalline solids, it is possible to construct a complete set of solutions of the 
Schr\"odinger equation (the Floquet states) which are periodic in time up to a phase
\begin{equation}
  \ket{\Psi_\alpha\left(t\right)} = \nep^{-i{\overline \mu}_\alpha t}\ket{\Phi_\alpha\left(t\right)} \;.
\end{equation}
The states $\ket{\Phi_\alpha(t)}$, the so-called Floquet modes, are periodic, $\ket{\Phi_\alpha(t+\tau)}=\ket{\Phi_\alpha(t)}$ while the real quantities 
${\overline \mu}_\alpha$ are called Floquet quasienergies.
%
If we assume that the system starts in the density matrix $\hat{\rho}_{0}$, we can expand the density matrix at time $t$ in the Floquet basis. 
Exploiting the fact that the Floquet states are solutions of the Schr\"odinger equation we see that 
$\bra{\Psi_\alpha (t)} \hat{\rho}(t) \ket{\Psi_\beta(t)} = \bra{\Phi_\alpha(0)} \hat{\rho}_0 \ket{\Phi_\beta(0)}$. 
Defining $\rho_{\alpha \beta}(0) \equiv \bra{\Phi_\alpha(0)} \hat{\rho}_0 \ket{\Phi_\beta(0)}$ we can write
\begin{equation}
  \hat{\rho}(t) = \sum_{\alpha \beta} \nep^{-i\left({\overline \mu}_\alpha-{\overline \mu}_\beta\right)t} \rho_{\alpha \beta}(0) \ket{\Phi_\alpha(t)} \bra{\Phi_\beta(t)} \;.
\end{equation}
The mean value of the operator $\hat{A}$ at time $t$ is therefore
\begin{equation}
  \mean{A}_t = \Tr[  \hat{\rho}(t) \hat{A} ] = \sum_{\alpha \beta} \nep^{-i\left({\overline \mu}_\alpha-{\overline \mu}_\beta\right)t}  \rho_{\alpha \beta}(0) A_{\beta\alpha}(t) \;,
\end{equation}
where we have defined $A_{\beta\alpha}(t)=\bra{\Phi_\beta(t)} \hat{A} \ket{\Phi_\alpha(t)}$, 
which is, by construction, a $\tau$-periodic quantity. 
We can then divide the previous sum into two parts: a periodic one, originating from diagonal elements, 
and an extra piece, originating from off-diagonal elements:
\begin{equation} \label{averator}
  \mean{A}_t = \mean{A}_t^{\rm diag}+\mean{A}_t^{\rm off-diag} \;.
\end{equation}
We can express these two contributions, assuming a non-degenerate Floquet spectrum (${\overline \mu}_\beta\neq {\overline \mu}_\alpha$ if $\beta\neq \alpha$), as follows:
\footnote{If strict degeneracies are present, the periodic part would get contributions from off-diagonal terms with ${\overline \mu}_\beta={\overline \mu}_\alpha$.
See Sec.~\ref{subchain:sec} for a discussion of \emph{quasi}-degenracies tending to strict degeneracies in the thermodynamic limit in the case of a local perturbation.}
\begin{eqnarray} 
\label{aver-per}
\mean{A}_t^{\rm diag} &\equiv& \sum_{\alpha} \rho_{\alpha \alpha}(0) A_{\alpha\alpha} (t)  \\
\label{aver-fluc}
\mean{A}_t^{\rm off-diag} &\equiv& \int_{-\infty}^{+\infty} \! \frac{\ud\omega}{\pi} \;\; F_t(\omega) \; \nep^{-i\omega t} \;,
\end{eqnarray}
where we have introduced the time-dependent $\tau$-periodic weighted joint density of states
\begin{equation} \label{outofdiag}
  F_t (\omega) \equiv \pi \sum_{\alpha\neq\beta} \rho_{\alpha\beta}(0) A_{\beta \alpha} (t) \delta\left( \omega-{\overline \mu}_\alpha+{\overline \mu}_\beta\right) \;.
\end{equation}
Suppose we now evaluate $\mean{A}_t^{\rm off-diag}$ at an arbitrary time $t_0 + n\tau$, where $t_0\in [0,\tau]$. 
Since $F_{t_0+n\tau} (\omega) = F_{t_0} (\omega)$, we can readily find that:
\[
\mean{A}_{t_0+n\tau}^{\rm off-diag} = \int_{-\infty}^{+\infty} \! \frac{\ud\omega}{\pi} \;\; F_{t_0}(\omega) \; \nep^{-i\omega (t_0+n\tau)} \;,
\]
i.e., exactly of the form to which the Riemann-Lebesgue lemma, Eq.~\eqref{RL:eqn}, might apply.
If $F_t (\omega)$ is a sufficiently smooth function of $\omega$ (such that $|F_t(\omega)|$ is Lebesgue-integrable) one would conclude that 
$\mean{A}_t^{\rm off-diag} $ decays to 0 after a transient, and the resulting large-$t$ behaviour of $\mean{A}_t$ is asymptotically periodic,
$\mean{A}_t \longrightarrow \mean{A}_t^{\rm diag}$.
As discussed in \cite{Russomanno_PRL12},  this occurs whenever the Floquet spectrum is a continuum 
(in the absence of singularities).
This vanishing of the fluctuating piece, and the resulting time-periodic response, will be henceforth referred to as ``synchronization'' \cite{Russomanno_PRL12}.
As we will argue later, this off-diagonal term appears to acquire a singular contribution whenever the driving is local, thus leading to a steady
energy absorption which, however, is not extensive (see Sec.~\ref{energy:sec} and Sec.~\ref{subchain:sec} for a discussion of this point).
%
%
%

\section{Energetic considerations: synchronization versus absorption} \label{energy:sec}
%
In the following, we will discuss the physics of energy absorbtion in a system described by the generic Hamiltonian 
$\hat{H}(t)=\hat{H}_0+v(t)\hat{A}$. For this sake, it is convenient to define two energy functions: 
the first, $E_0(t) = \Tr [ \hat{\rho}(t) \hat{H}_0 ]$, is the energy of the original system in the perturbed state $\hat{\rho}(t)$, while
the other, $E(t) = \Tr [ \hat{\rho}(t) \hat{H}(t) ]= E_0(t) + v(t) \mean{A}_t $, is the total energy including the perturbing-field term.
Using the fact that a coherent unitary evolution implies
$i\hbar \dot{\hat{\rho}}(t) =   [\hat{H}(t),\hat{\rho}(t)]$, together with
%
%
the cyclic property of the trace, it is easy to derive a Hellmann-Feynmann-like formula:
\begin{equation} \label{Et:eqn}
\frac{d}{dt} E(t) = \Tr[  \hat{\rho}(t) \frac{d}{dt} \hat{H}(t) ] = \dot{v}(t) \mean{A}_t \;.
\end{equation}
On the other hand, using  $E_0(t) = E(t) - v(t) \mean{A}_t $, taking a derivative,  
it is straightforward to conclude that:
\begin{equation} \label{E0t:eqn}
\frac{d}{dt} E_0(t) = -v(t) \frac{d}{dt} \mean{A}_t \;.
\end{equation}
Consider now for definiteness, $v(t)=v_0\sin{(\omega_0 t)}$ as in Section \ref{linear-response:sec}.
The energy change during the $n$th oscillation of the field, i.e., in the time window $[(n-1)\tau,n\tau]$, is given by
$\Delta E(n) = E(n\tau)-E((n-1)\tau)$ and $\Delta E_0(n) = E_0(n\tau)-E_0((n-1)\tau)$. Both are directly obtained
from Eqs.~(\ref{Et:eqn}-\ref{E0t:eqn}) (the second, through integration by parts) and have the form:
\begin{equation} \label{DeltaE_n:eqn} 
\Delta E(n) = \Delta E_0(n) = v_0 \omega_0 \int_{(n-1)\tau}^{n\tau} \! dt \; \cos{(\omega_0 t)} \mean{A}_t \;.
\end{equation}
If we consider the restriction of $\mean{A}_t$ to the $n$th-period time-window  $[(n-1)\tau,n\tau]$, call it 
$\left[\mean{A}_t \right]_n$, we can expand it in a standard Fourier series
\begin{equation} 
  \left[\mean{A}_t \right]_n =\widetilde{{A}}_0(n) + \sum_{m=1}^{+\infty} \left[ \widetilde{A}_m^{(c)}(n) \cos (m\omega_0 t)+
                                                                                                                           \widetilde{A}_m^{(s)}(n) \sin (m\omega_0 t) \right] \,,
\end{equation}
where the Fourier coefficients $\widetilde{A}_m^{(c,s)}$ depend in general on the time-window index $n$,
because the off-diagonal piece $\mean{A}_t^{\rm off-diag}$ makes $\mean{A}_t= \mean{A}_t^{\rm diag}+\mean{A}_t^{\rm off-diag}$ to be not strictly periodic. 
Evidently, see Eq.~\eqref{DeltaE_n:eqn}, the coefficient $\widetilde{A}_1^{(c)}(n)$ of the $\cos{(\omega_0t)}$ component is what determines
the rate of energy absorption: 
\begin{equation} \label{raten:eqn}
{\cal W}_n = \frac{\Delta E(n)}{\tau} =   \frac{1}{2} v_0 \omega_0  \widetilde{A}_1^{(c)}(n) \;.
\end{equation}
LRT  predicts in the steady state (after the decay of the transient Eq.~\eqref{LR-transient:eqn}), an out-of-phase response with $\widetilde{A}_1^{(c)}=-v_0 \chi''(\omega_0)$, 
which leads to a steady-state ($n\to \infty$) increase of the energy at a rate \cite{Giuliani-Vignale:book}
\begin{equation} \label{W:eqn}
  \mathcal{W}^{\rm LRT}_{n\to \infty} = -\frac{1}{2} \omega_0 v_0^2 \chi''(\omega_0) > 0 \;,
\end{equation}
which is positive, since $\chi''(\omega_0>0)<0$.

As mentioned in the Introduction, it is well known in the context of mesoscopic physics that whenever a system is closed, energy absorption resulting from an oscillatory perturbation 
(for example in disordered mesoscopic rings)
does not correspond to energy dissipation, but rather energy storage~\cite{Landauer_PRB86}. 
Most importantly, in these systems the energy absorption rate, which classically would be constant, tends to decrease at long times as a result of dynamical localization~\cite{Gefen_PRL87}.
A steady increase of energy is problematic not only for mesoscopic systems but also for closed system on a  lattice, say a fermionic Hubbard-like model, 
the transverse field quantum Ising model, or any spin model in any dimension, whenever $\hat{A}$ is an extensive operator. 
Indeed, by simple arguments one can show that the spectrum of such Hamiltonians on a lattice of $N=L^D$ sites, $D$
being the dimensionality of the lattice, should be bounded in a region $[e_L N,e_U N]$, where $e_L$ and $e_U$
are appropriate finite lower and upper bounds on the energy-per-site. 
If $\hat{A}$ is extensive, then a steady ($n \to \infty$) and extensive ($\propto N$) energy increase with a rate $\mathcal{W}$, like that predicted by LRT, 
would inevitably lead to a violation of the boundedness of the spectrum: $|E(t)-E(0)|/N < |e_U-e_L|$. 
Local operators, on the contrary, do not lead to an extensive energy increase, and do not violate any bound.
We therefore expect, and explicitly illustrate in the following, that LRT should eventually break down after a while when the perturbation is extensive, even
if we are in the thermodynamic limit.

%

As in the case of mesoscopic systems, also in the case of closed system on a lattice energy absorption can be hindered. 
In particular, this happens  if all the observables ``synchronize'' with the perturbing field \cite{Russomanno_PRL12} by
showing, after a transient, a perfectly periodic asymptotic response. When this happens $\mean{A}_t \to \mean{A}_t^{\rm diag}$
and $\widetilde{A}_1^{(c)}(n)\to 0$, for large $n$, making the energy absorption rate vanish at large times. 
In this case the response is asymptotically ``in-phase'' with the perturbation: 
%
%
\begin{equation}
\mean{A}_t \longrightarrow \mean{A}_t^{\rm diag} =  \widetilde{{A}}_0 +  \widetilde{A}_1^{(s)} \sin (\omega_0 t) + (\mbox{higher harmonics}) \;, 
\end{equation}
{\em without the out-of-phase term} proportional to $\cos{(\omega_0 t)}$. 
%
\begin{figure}
\begin{center}
   \includegraphics[width=9cm]{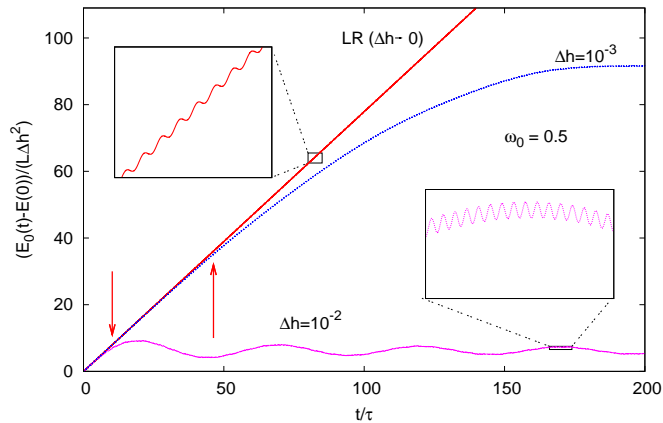}
\end{center}
\caption{Plot of the energy absorbed per site, vs $t/\tau$, for $\omega_0=0.5$, for a one dimensional transverse field Ising model which is perturbed, 
around the critical point, with a uniform transverse field modulation $(\Delta h) \sin{(\omega_0 t)}$. Details expained in Section \ref{Ising:sec}.
The red solid line is the LRT result, compared to the exact results for $\Delta h = 10^{-2}$ (purple dotted line) and $\Delta h = 10^{-3}$ (blue dashed line). 
All the results are rescaled by $1/(\Delta h)^2$ so as to make the comparison meaningful.
The limit $\Delta h\to 0$ is evidently singular.}
\label{workers:fig}
\end{figure}
%
Figure \ref{workers:fig} illustrates this with an explicit calculation performed on the one-dimensional transverse field Ising model, 
whose results will be detailed in Section \ref{Ising:sec}. 
The solid red line represents the energy absorbed per site $(E_0(t)-E_0(0))/L$ versus the rescaled time $t/\tau$ within LRT, for an Ising chain whose transverse
field is uniformly modulated, around the critical point $h_c=1$, by a term $(\Delta h) \sin{(\omega_0 t)}$ with $\omega_0=0.5$, corresponding to a part of the spectrum where $\chi''(\omega_0)\neq 0$.
Observe the overall linear increase in time with a positive average rate of absorption $\mathcal{W}$, Eq.~\eqref{W:eqn}, with superimposed small oscillations on the scale of the
period $\tau$. The other two lines represent the corresponding exact results, obtained from a Floquet analysis, for $\Delta h=10^{-3}$ and $\Delta h=10^{-2}$  
(all results have been rescaled by $1/(\Delta h)^2$ to make the comparison meaningful). 
We observe that for any small but finite $\Delta h$ the exact results eventually deviate, for large $t$, from the linear-in-$t$ LRT prediction, saturating at large times up to small and larger-scale oscillations.
%
Similar physics apparently emerges, for instance, in a periodically modulated homogeneous one-dimensional Hubbard model \cite{Kollath_PRA06},
as numerically found through time-dependent density matrix renormalization group (t-DMRG) calculations \cite{Daley_JSTAT04,White_PRL04}, 
see Fig. 1 of Ref.~\cite{Kollath_PRA06}.
The possible experimental relevance of this departure, provided the time-scale for coherent evolution is large enough, is discussed in Section \ref{conclusions:sec}.
We mention that there are other possible scenarios by which a system can stop absorbing energy indefinitely: one involves a $\widetilde{A}_1^{(c)}(n)\to 0$ but without a full vanishing 
of $\mean{A}_t^{\rm off-diag}$, another a $\widetilde{A}_1^{(c)}(n)$ that keeps oscillating around $0$ in such a way that $(E(t)-E(0))/N$ remains bounded.

\section{Quantum Ising chain under periodic transverse field} \label{Ising:sec}
%
In this section we corroborate the previous arguments with detailed calculations of the dynamics of a quantum Ising chain in transverse field. 
After introducing the model, we discuss first the LRT approximation, then the exact Floquet analysis, and finally we compare the two results.
The Hamiltonian of the system is
%
%
\begin{equation}  \label{h11}
  \hat{H}(t) =-\frac{1}{2}\sum_{j=1}^{L}\left(J\sigma_j^z\sigma_{j+1}^z + h \sigma_j^x\right)  + v(t) \sum_{j=1}^{l} \sigma_j^x  \;.
\end{equation}
Here, the $\sigma^{x,z}_j$ are spins (Pauli matrices) at site $j$ of a chain of length $L$ with periodic boundary conditions $\sigma^{x,z}_{L+1}=\sigma^{x,z}_1$,  
$J$ is a longitudinal coupling ($J=1$ in the following), while the transverse field has a uniform piece, $h$, and a time-dependent one, $\propto v(t)$, acting only on a
subchain of length $l$. In the following we will take $v(t)$ to be periodic, parameterizing it as $v(t)=-(\Delta h/2) \theta(t) \sin(\omega_0 t)$. 
%
This Hamiltonian can be transformed, through a Jordan-Wigner transformation \cite{Lieb_AP61}, to a ``solvable'' quadratic-fermion form. 
At equilibrium and for a homogenous transverse field, $v(t)=0$, the model has two mutually dual gapped phases, 
a ferromagnetic ($|h|<1$),  and a quantum paramagnetic ($|h|>1$), separated by a quantum phase transition at $h_c=1$.
When $\Delta h>0$, the transverse field starts oscillating periodically, for $t\ge 0$ and in a region of size $l$, around the uniform value $h$.
In the notation of Sec.~\ref{linear-response:sec}, $\hat{H}(t)=\hat{H}_0 + v(t) \hat{A}$ where $\hat{H}_0$ is the homogeneous model with transverse field $h$ 
(which we will set for convenience to critical value $h=h_c=1$) and $\hat{A}=\hat{M}_l=\sum_{j=1}^l \sigma_j^x$ is the transverse magnetization of a region comprising $l$ sites.
We start discussing the extensive case with $l=L$ (the periodic driving acts on the whole chain), where translational invariance
simplifies the analysis considerably, since $\hat{A}=\hat{M}_L=\sum_{j=1}^L \sigma_j^x$. 
Further technical details for the general non-translationally invariant case are contained in \ref{Bogoliubov:sec}.
When $l=L$, going to $k$-space, $\hat{H}(t)$ becomes a sum of two-level systems: 
\begin{equation} \label{Ht:eqn}
  \hat{H}(t) = \sum_k^{\rm ABC} \hat{H}_k(t) = \sum_k^{\rm ABC}
                    \left(\begin{array}{cc}
			c_k^\dagger & c_{-k}
		\end{array}\right)
	\left(\begin{array}{cc}
			E_k(t) & -i\Delta_k\\
			\\
			i\Delta_k & -E_k(t)
		\end{array}\right)
	\left(\begin{array}{c}
			c_k \\
			\\
			c_{-k}^\dagger
		\end{array}\right) \;, 
\end{equation}
where $E_k(t) = h(t)-\cos{k}$, $\Delta_k=\sin{k}$, and the sum over $k$ is restricted to positive $k$'s of the form $k=(2n+1)\pi/L$ with $n=0,\ldots,L/2-1$,
corresponding to anti-periodic boundary conditions (ABC) for the fermions \cite{Lieb_AP61}, as appropriate for $L$ multiple of $4$, which we assume.
We will briefly refer to such a set of $k$, in the following, as $k\in {\rm ABC}$.
Each $\hat{H}_k(t)$ acts on a 2-dim Hilbert space generated by $\{ c_k^\dagger c_{-k}^\dagger \ket{0}, \ket{0} \}$,
and can be represented in that basis by a $2\times 2$ matrix
$H_k(t)=E_k(t) \sigma^z + \Delta_k \sigma^y$, with instantaneous eigenvalues 
$\pm\sqrt{E_k^2(t)+\Delta_k^2}$.
In the same representation, the unperturbed (critical) Hamiltonian is given by
\begin{equation} \label{H0:eqn}
\hat{H}_0 = \sum_k^{\textrm{ABC}} \hat{H}_k^0 = \sum_k^{\textrm{ABC}} 
                  \left(\begin{array}{cc}
			c_k^\dagger & c_{-k}
		\end{array}\right)
	\left(\begin{array}{cc}
			1-\cos(k) & -i\sin(k)\\
			\\
			i\sin(k) & \cos(k)-1
		\end{array}\right)
	\left(\begin{array}{c}
			c_k \\
			\\
			c_{-k}^\dagger
		\end{array}\right)\;,	
\end{equation}
with eigenvalues given by  $\pm\epsilon_k^0 = \pm 2\sin(k/2)$.
This immediately implies that the natural resonance frequencies are at $\pm 2\epsilon_k^0$, which in our units are between $-4$ and $4$.

We assume that the coherent evolution starts with the system in the ground state at time $0$, which has the BCS-like form
\begin{equation} \label{ground}
	\ket{\Psi_{\rm GS}} = \prod_{k>0}^{\rm ABC} \ket{\psi_k^0} = \prod_{k>0}^{\rm ABC} \left(u_k^{0}+v_k^{0}c_k^\dagger c_{-k}^\dagger\right) \ket{0} \;,
\end{equation}
with $u_k^{0}=\cos(\theta_k/2)$ and $v_k^{0}=i\sin(\theta_k/2)$ obtained by diagonalizing the $2\times 2$ problem 
in Eq.~\eqref{H0:eqn} in terms of the angle $\theta_k$, given by $\tan{\theta_k} = (\sin{k})/(1-\cos{k})$.
For future reference, we mention that the equilibrium (ground state) value of the transverse magnetization density is given by
$m_{\rm eq} \equiv \bra{\Psi_{\rm GS}} \hat{m} \ket{\Psi_{\rm GS}}$, which in the thermodynamic limit equals $m_{\rm eq}=2/\pi$.

\subsection{Linear response theory approximation for $l=L$}
%
The time-dependent  modulation of the transverse field present in $\hat{H}(t)$ is given by $-\theta(t)(\Delta h/2) \sin(\omega_0 t) \hat{M}_L$.
In the notation of Section \ref{linear-response:sec}, this implies a $v_0=-\Delta h/2$ and $\hat{A}=\hat{M}_L$. 
In order to have a meaningful thermodynamic limit, we calculate the zero-temperature perturbed value of the transverse magnetization density 
$\hat{m}=\hat{M}_L/L$ by the corresponding susceptibility:
\begin{equation} \label{chitty}
  \chi(t) \equiv-\frac{i}{\hbar}\,\theta(t) \bra{\Psi_{\rm GS}}\left[\hat{m}(t),\hat{M}_L\right]\ket{\Psi_{\rm GS}} \;.
\end{equation}
As shown in \ref{trasmag:app}, the corresponding $\chi''(\omega_0)$ is given, in the thermodynamic limit $L\to \infty$, by
\begin{equation} \label{chi-ising:eqn}
   \chi''(\omega_0) = - {\rm sign}(\omega_0)  \, \theta(4-|\omega_0|) \, \sqrt{1-\left(\frac{\omega_0}{4}\right)^2} \;.
\end{equation}
We notice that $\chi''(\omega_0)$ is odd and nonvanishing only provided $\left| \omega_0 \right|<4$, 
that is when the driving frequency falls inside the spectrum of the natural resonance frequencies of the system.
The corresponding $\chi'(\omega_0)$ is calculated using  Eq.~\eqref{chi-prime:eqn}.
The functions $\chi'(\omega_0)$ and $\chi''(\omega_0)$ will be shown in Fig.~\ref{chi_mo_om:fig}. 
%
Summarizing, the LRT prediction for the transverse magnetization density is:
\begin{equation} \label{levolotion}
  m_{\rm LRT}^{\rm per} (t) = m_{\rm eq}  - \frac{\Delta h}{2} \left[ \chi'(\omega_0) \sin\left(\omega_0 t\right) - \chi''(\omega_0) \cos\left(\omega_0 t\right) \right] + F^{\rm trans}(\omega_0,t) \;,
\end{equation}
where the transient part $F^{\rm trans}(\omega_0,t)$ is given by Eq.~\eqref{LR-transient-2:eqn} with $v_0=-(\Delta h)/2$.

\subsection{Exact evolution and Floquet theory for $l=L$} \label{floquis}
%
We describe here the exact evolution of the magnetization expressed through a Floquet analysis \cite{Russomanno_PRL12},
as an exemplification of the general arguments of Section \ref{floquet:sec}. 
Details on how to compute Floquet modes and quasienergies in this case are given in \cite{Russomanno_PRL12} and the related supplementary material. There,
and in~\ref{Bogoliubov:sec}, we explain also how to extend this picture to the non-uniform case; in this section we focus on the uniform one
because it is more transparent and instructive.

The state of the system at all times can be written in a BCS form
\begin{equation}
  \ket{\Psi(t)}=\prod_{k>0}^{\rm ABC} \ket{\psi_k(t)}=\prod_{k>0}^{\rm ABC} \left(u_k(t)+v_k(t)c_k^\dagger c_{-k}^\dagger \right) \ket{0} \;,
\end{equation}
where the functions $u_k(t)$ and $v_k(t)$ must obey the Bogoliubov-De Gennes equations 
\begin{equation} \label{deGennes:eqn}
    i\hbar \frac{d}{dt}\left(\begin{array}{cc}
			v_k(t)\\u_{k}(t)
           		\end{array} \right)
		= \left(\begin{array}{cc}
			\epsilon_k(t) &-i\Delta_k\\
			i\Delta_k&-\epsilon_k(t)
		\end{array} \right)
		\left(\begin{array}{cc}
			v_k(t)\\u_{k}(t)
		\end{array} \right) 
	\;,
\end{equation}
with initial values $v_k(0)=v_k^{0}$ and $u_k(0)=u_k^{0}$, because at time $t=0$ the system is in the ground state \eqref{ground}. 
The dynamics is quite clearly factorized in the two-dimensional subspaces generated by $\{ c_k^\dagger c_{-k}^\dagger \ket{0}, \ket{0} \}$.

The transverse magnetization operator $\hat{M}_L$ reads, in terms of Jordan-Wigner fermions, as 
$\hat{M}_L=\sum_{k>0}^{\textrm{ABC}} \hat{m}_k$ where $\hat{m}_k = 2 \left( c_{-k} c_{-k}^\dagger - c_k^\dagger c_k \right)$.
Using this, we can express the average transverse magnetization density at time $t$, in the thermodynamic limit, as:
\begin{equation} \label{magnets}
  m(t) = \int_0^\pi \! \frac{\ud k}{2\pi} \; \bra{\psi_k(t)} \hat{m}_k \ket{\psi_k(t)} \;.
\end{equation}
In each $k$-subspace, the state can be expanded in the Floquet basis
\begin{equation}
  \ket{\psi_k(t)} = r_k^+ \nep^{-i\mu_k t} \ket{\phi_k^+(t)} + r_k^- \nep^{i\mu_k t} \ket{\phi_k^-(t)} \;,
\end{equation}
where $r_k^{\pm} = \left\langle \phi_k^{\pm}(0) \right. \ket{\psi_k(0)}$ are the overlap factors between the 
initial state $\ket{\psi_k(0)}$ and the Floquet modes $\ket{\phi_k^{\pm}(t)}$ with Floquet quasi-energies $\pm\mu_k$  
(the quasi-energies have an opposite sign because the Hamiltonian Eq.~\eqref{Ht:eqn} has a vanishing trace). 
Substituting this in Eq.~\eqref{magnets} and separating diagonal and off-diagonal matrix elements, in strict analogy with what done in Section \ref{floquet:sec}, 
we can write $m(t)$ as a sum of two contributions, a $\tau$-periodic and a fluctuating one
\begin{equation}
m(t) = m^{\rm diag}(t) + m^{\rm off-diag}(t) \;,
\end{equation}
where:
\begin{eqnarray} 
\label{mper:eqn}
  m^{\rm diag}(t) &=&  \sum_{\alpha=\pm} \int_0^\pi \! \frac{\ud k}{2\pi} \; \left| r_k^\alpha \right|^2 \bra{\phi_k^\alpha(t)} \hat{m}_k \ket{\phi_k^\alpha(t)} \\
\label{mfluc:eqn}
  m^{\rm off-diag}(t) &=& \int_0^\pi \! \frac{\ud k}{\pi} \; \Real\left({r_k^+}^*r_k^- \bra{\phi_k^+(t)} \hat{m}_k \ket{\phi_k^-(t)} \nep^{-2i\mu_k t} \right) \;.
\end{eqnarray}
These expressions are the strict analogues of Eqs.~\eqref{averator}-\eqref{aver-fluc}: $m^{\rm diag}(t)$ is periodic in time,
 while $m^{\rm off-diag}(t)$ vanishes after a transient 
due to the Riemann-Lebesgue lemma, since the $\mu_k$, the overlaps $r_{k}^{\pm}$ and the matrix element $\bra{\phi_k^+(t)} \hat{m}_k \ket{\phi_k^-(t)}$ are continuous 
functions of $k$ (see discussion below, and Figure \ref{incontri:fig})). 
This result, first derived in \cite{Russomanno_PRL12}, implies that, after a transient, the transverse magnetization reaches a periodic ``steady regime''.
How long is the transient, depends on $\Delta h$: the smaller is $\Delta h$ the longer is the transient, until the singularities emerging for $\Delta h\to 0$ make
$m^{\rm off-diag}(t)$ no longer decaying to $0$.  
%
%

\subsection{Comparison of LRT against exact results for $l=L$} \label{comparison_l=L:sec}
%
Let us now discuss the results of an exact analysis in the regime of small $\Delta h$, where LRT should apply.
As already observed in Fig.~\ref{workers:fig}, LRT gives a good description of the energy absorbed at short times. 
%
\begin{figure}
\begin{center}
   \includegraphics[width=9cm]{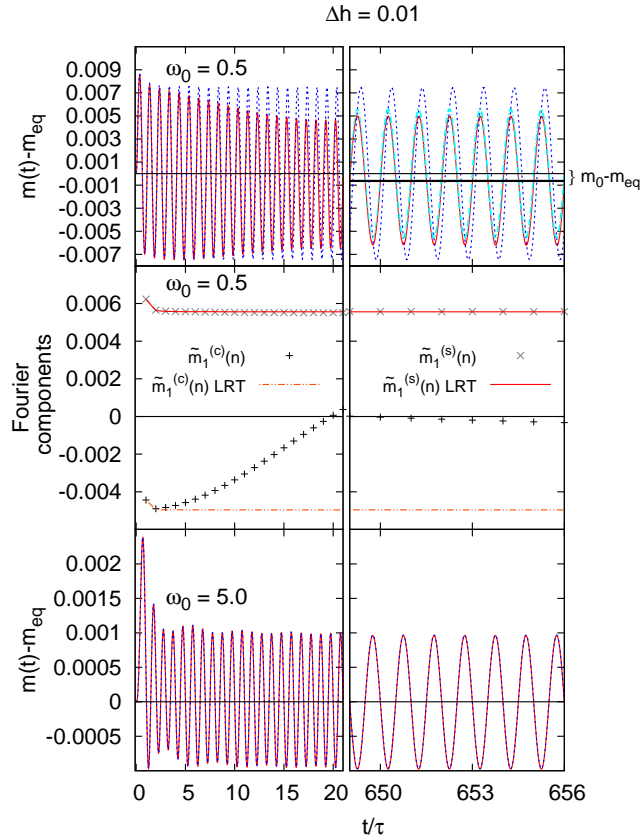}
\end{center}
\caption{
Plot of the exact transverse magnetization per spin $m(t)-m_{\rm eq}$ (red solid line) versus $t$ for a small driving field amplitude $\Delta h=10^{-2}$,
compared to the LRT prediction (blue dotted line). The upper panels are for $\omega_0=0.5$, where $\chi''(\omega_0)\neq 0$; 
the lower ones for $\omega_0=5$, where $\chi''(\omega_0)=0$. 
The upper right panel shows the disagreement between the exact $m(t)$ and LRT:
the exact $m(t)$ lacks the out-of-phase term proportional to $\chi''(\omega_0)$ and is slightly shifted downwards. The exact value 
is, on the contrary, well reproduced, a part a small downwards shift, by the in-phase term (proportional to $\chi'(\omega_0)$) (light blue dashed line).
The two central panels represent the Fourier coefficients $\widetilde{m}_1^{(c,s)}(n)$ of the $\cos{(\omega_0t)}$ and $\sin{(\omega_0t)}$
components of $m(t)$ in the time-window $[(n-1)\tau,n\tau]$ for $\omega_0=0.5$. While the latter tends, as expected, to the LRT counterpart for $n\to\infty$,
the former tends to 0.
}
\label{magnetar:fig}
\end{figure}
%
Fig.~\ref{magnetar:fig} shows the exact $m(t)-m_{\rm eq}$ versus $t$ (solid line), compared to the LRT result (dashed line), 
for $\Delta h = 10^{-2}$ and  two values of $\omega_0$: $\omega_0=0.5$ (upper panels),  where $\chi''(\omega_0)\ne 0$, 
and $\omega_0=5$ (lower panels),  where $\chi''(\omega_0)=0$.
The agreement is perfect in the first few periods of the driving (left panels), where we clearly see the effect of a transient even in LRT.
For larger $t$, the agreement is still perfect when $\omega_0=5$ (lower right panel), while it is evidently lost for $\omega_0=0.5$ (upper right panel). 
The upper right panel of Fig.~\ref{magnetar:fig}, in particular, deserves a few extra comments. 
The true response is evidently out-of-phase with respect to the prediction of LRT. Indeed, we observe that $m(t)-m_{\rm eq}$ is
essentially given by the in-phase LRT result $-(\Delta h/2) \chi'(\omega_0) \sin{(\omega_0 t)}$ (shown by a dashed-dotted line), apart for a small shift downwards:
in other words, $m(t)$ oscillates in phase with the perturbing field, but around an average value $\widetilde{m}_0<m_{\rm eq}$.
Summarizing we find that, for $\Delta h$ of order $10^{-2}$ or smaller, the large-$t$ behaviour of the exact $m(t)$ is
given by 
\begin{equation} \label{mper-t-Fourier:eqn}
  m(t) \stackrel{\scriptscriptstyle t\to \infty} {\longrightarrow} m^{\rm diag}(t) = \widetilde{m}_0 + \widetilde{m}_1^{(s)} \sin(\omega_0 t) \; + (\cdots ) \;,
\end{equation}
where $(\cdots )$ denote higher harmonics, whose Fourier coefficients we find to be of order $(\Delta h)^2$ or smaller.
As detailed in the central panels of Fig.~\ref{magnetar:fig}, the Fourier coefficient $\widetilde{m}_1^{(s)}(n)$ is correctly given by LRT, 
and quickly reaches the asymptotic value $\widetilde{m}_1^{(s)} = - (\Delta h/2) \chi'(\omega_0) + o(\Delta h)$, 
while the Fourier coefficient  $\widetilde{m}_1^{(c)}(n)$ of the out-of-phase term $\cos(\omega_0 t)$ --- which LRT predicts to be  
$(\Delta h/2) \chi''(\omega_0)$ --- rapidly drops to a value which decays (with oscillations) towards zero, $\widetilde{m}_1^{(c)}(n\to \infty)=0$, 
in agreement with the considerations of Section \ref{energy:sec} (the steady regime response $m^{\rm diag}(t)$ is syncronized in-phase with the driving).
Moreover, the zero-frequency Fourier coefficient $\widetilde{m}_0$ differs from $m_{\rm eq}$ by terms of linear order in $\Delta h$ when $\chi''(\omega_0)\ne 0$.
These results, which we have verified for all frequencies $\omega_0$ are summarized in Fig.~\ref{chi_mo_om:fig}.
%
\begin{figure}
\begin{center}
   \includegraphics[width=9cm]{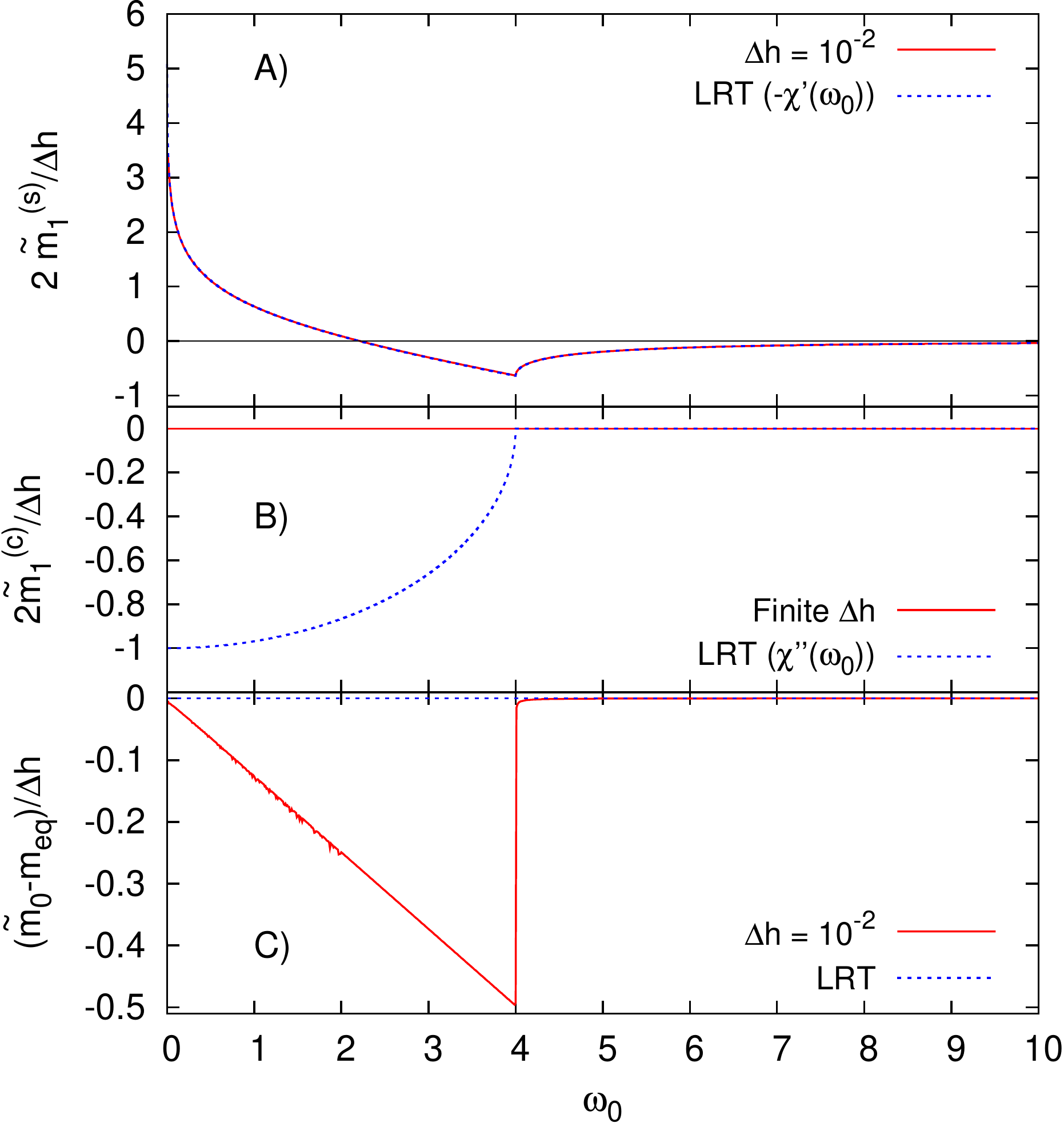}
\end{center}
\caption{(A) Plot of $2\widetilde{m}_1^{(s)}/(\Delta h)$, where  $\widetilde{m}_1^{(s)}$ is the Fourier coefficient of the $\sin(\omega_0 t)$ 
in Eq.~\eqref{mper-t-Fourier:eqn}, versus  $\omega_0$, compared with the LRT prediction $-\chi'(\omega_0)$.  
(B)  A similar plot for $2\widetilde{m}_1^{(c)}/(\Delta h)$, where  $\widetilde{m}_1^{(c)}$ is the Fourier coefficient of the $\cos(\omega_0 t)$ term, 
which vanishes exactly, while it is predicted to be $\chi''(\omega_0)$ within LRT.
(C) Plot of $(\widetilde{m}_0-m_{\rm eq})/(\Delta h)$, where $\widetilde{m}_0$ is the zero-frequency (i.e., constant) Fourier coefficient in 
Eq.~\eqref{mper-t-Fourier:eqn}.  In all panels, the exact Fourier coefficients are shown with red solid lines, the corresponding LRT by blue dotted lines.}
\label{chi_mo_om:fig}
\end{figure}
%

\begin{figure}
\begin{center}
   \includegraphics[width=9cm]{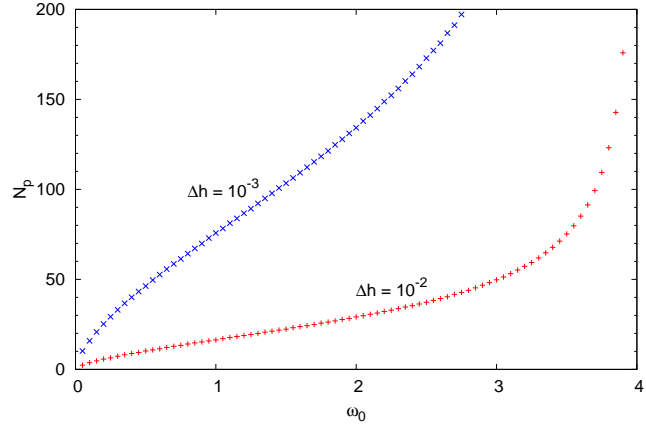}
\end{center}
\caption{Plot of the approximate number of periods $t^*/\tau$ over which LRT is accurate for a uniformly driven quantum Ising chain 
as a function of the frequency $\omega_0$, for two values of $\Delta h$ .}
\label{periodi:fig}
\end{figure}

Let us go back to the issue of energy absorption. 
As discussed in Section \ref{energy:sec}, the out-of-phase term, proportional to $\chi''(\omega_0)$, appearing within LRT (see Eq.~\eqref{levolotion}) results in a net
energy absorption for large $t$ with a constant rate $\mathcal{W}$ (see Eq.~\eqref{W:eqn}). This large-$t$ steady absorption is absent in the true response: Eq.~\eqref{raten:eqn} implies that the energy absorption rate tends asymptotically to 0 together with the cosine component $\widetilde{m}_1^{(c)}(n)$ plotted
 in Fig.~\ref{magnetar:fig}. 
This is better seen in Figure \ref{workers:fig}, which illustrates the energy-per-site absorbed at time $t$, $(E_0(t) - E_0(0))/L$,
for $\omega_0=0.5$ and two values of $\Delta h$: $\Delta h=10^{-2}$ and $\Delta h=10^{-3}$. 
As discussed above, the LRT prediction grows with a rate $\mathcal{W}$ given precisely by Eq.~\eqref{W:eqn}, i.e., $\mathcal{W}/L= -(\omega_0/8)(\Delta h)^2 \chi''(\omega_0)$.
The arrows in Figure \ref{workers:fig} indicate the time $t^*$ at which the exact values of $(E_0(t) - E_0(0))/L$ differ from the LRT result by a quantity $(\Delta h)^2$.
This time $t^*$ is longer for decreasing values of $\Delta h$, and depends also on $\omega_0$. From similar data, one can extract information on
the approximate number of periods of the driving, $t^*/\tau$, for which LRT is accurate for various $\Delta h$ and $\omega_0$.   
This information is contained in Figure \ref{periodi:fig}. Notice that, especially in the low frequency region $\omega_0<1$, the number of periods
for which LRT works is remarkably small, of order of $10\div 60$ for $\Delta h=10^{-3}$ and down to numbers of order $1\div 10$, for $\Delta h=10^{-2}$
(for which, nominally, LRT is an excellent approximation, at least for what concerns the in-phase term).
%

To conclude this section, let us discuss how the principal-value singularities giving rise to the cosine term 
in LRT (see Eqs.~\eqref{meaner1} and \eqref{LR-transient:eqn}) become sharp but regular features when $\Delta h$ is small but finite, 
giving therefore rise to a vanishing transient (see Eqs.~\eqref{aver-fluc} and \eqref{outofdiag}) in the true evolution. 
The presence of a finite small $\Delta h$ provides a natural regularization for the principal-value singularities occuring in LRT.
To show this, consider again the out-of-phase contribution (o.o.p.) to the average $\delta m_{\rm LRT}(t)$ which is given by:
\footnote{Do the thermodynamic limit of Eq.~\ref{magliner} or put $\chi''(\omega)$ given by Eq.~\ref{chi-ising:eqn} in Eq.~\ref{meaner} 
and change the integration variable as $\omega=4\sin(k/2)$.} 
\begin{equation} \label{LRT-out:eqn}
m_{\rm LRT}^{\rm  o.o.p.}(t) = 2 \omega_0 \Delta h \; \dashint_0^\pi \! \frac{\ud k}{\pi} \; \frac{  \cos^2(k/2) }{ \omega_0^2 - (2\epsilon_k^0)^2} \sin(2\epsilon_k^0 t) \;.
\end{equation}
(The integration over $k$, as opposed to the integral over $\omega$, makes explicit the factorization
of the Hamiltonian into an ensemble of two-level subsystems labeled by $k$.)
This should be compared with the contribution to $m(t)$ originating from the off-diagonal elements in the Floquet expansion, given by
$m^{\rm off-diag}(t)$ in Eq.~\eqref{mfluc:eqn}, where the Floquet quasi-energies $\mu_k$ appear.
%
\begin{figure}
\begin{center}
   \includegraphics[width=9cm]{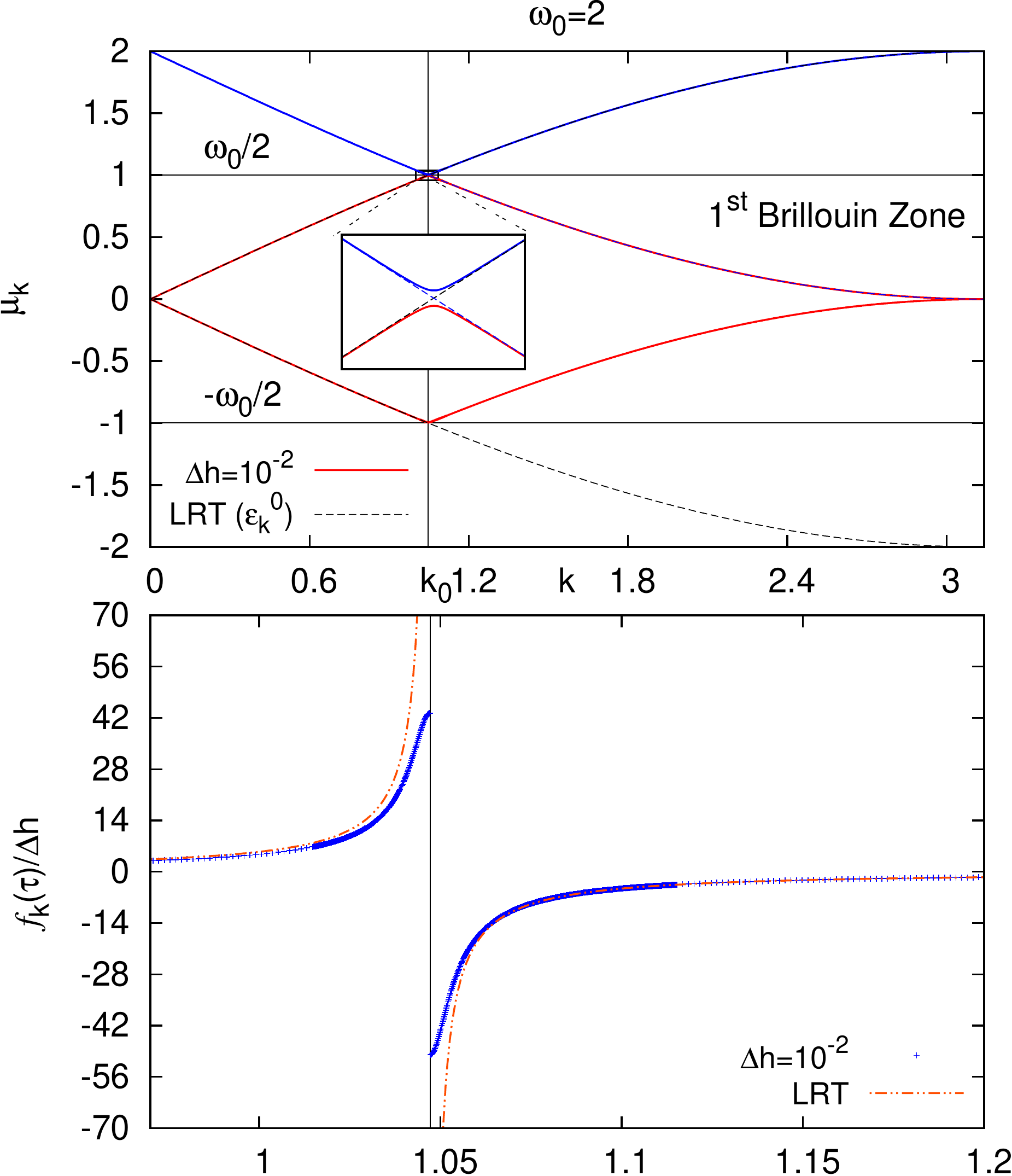}
\end{center}
\caption{(Top) The Floquet quasi-energies $\pm \mu_k$ (continuous lines) versus $k$ for a weak driving of $\Delta h =10^{-2}$ at $\omega_0=2$, compared to the
unperturbed excitation energies $\pm \epsilon_k^{0}$ (dashed lines). $\mu_k$ coincides with $\epsilon_k^0$ up to terms of order $(\Delta h)^2$ everywhere 
but around the (one-photon) resonance occurring at $2\epsilon_{k_0}^{0}=\omega_0$ (here $k_0=\pi/3$). 
The inset shows that the resonance is an avoided crossing of the Floquet exponents.
(Bottom) Plot of $f_k(\tau)$ (see Eq.~\eqref{fluct-app:eqn}) compared to the corresponding LRT diverging integrand $f_k^{\rm LRT}$ (see Eq.~\eqref{LRT-out:eqn})
close to the resonance point $k_0$, both rescaled by $\Delta h$. 
}
\label{incontri:fig}
\end{figure}
%
In the limit in which $\Delta h$ is small, $\mu_k$ approaches the unperturbed energy $\epsilon_k^0$ except at isolated 
resonance points. Figure \ref{incontri:fig} (upper panel) shows a plot of $\epsilon_k^0=2\sin{k/2}$ compared to $\mu_k$ for
$\omega_0=2$ and $\Delta h=10^{-2}$: notice the avoided crossing of $\mu_k$ at the border of the Floquet 
first Brillouin zone (1BZ)~\cite{Grifoni_PR98,Russomanno_PRL12} at $[-\omega_0/2,+\omega_0/2]$. 
In essence, when $\Delta h\to 0$, $\mu_k$ tends towards $\epsilon_k^0$, folded in the Floquet 1BZ. 
By taking due care of this folding, one can show that, for small $\Delta h$, the out-of-diagonal contribution $m^{\rm off-diag}(t)$
is approximately given by:
\begin{equation}  \label{fluct-app:eqn}
  m^{\rm off-diag}(t) \approx \int_0^\pi \! \frac{\ud k}{\pi} \;\left[g_k(t)\cos(2\epsilon_k^0 t) + f_k(t) \sin(2\epsilon_k^0 t)\right] \;,
\end{equation}
where the two $\tau$-periodic quantities $g_k(t)$ and $f_k(t)$ originate from the appropriate combinations of the
real and imaginary parts of the matrix element $F_k(t) =  {r_k^+}^* r_k^- \bra{\phi_k^+(t)} \hat{m}_k \ket{\phi_k^-(t)}$ 
appearing in Eq.~\eqref{mfluc:eqn}. Both $g_k(t)$ and $f_k(t)$ are regular functions with, at most, a discontinuity
across the resonance, while the corresponding LRT integrand $f_k^{\rm LRT}=2 \omega_0 \Delta h \cos^2(k/2)/ [\omega_0^2 - (2\epsilon_k^0)^2]$ 
is highly singular and requires a principal value prescription.
The lower part of Figure \ref{incontri:fig} shows the behaviour of $f_k(t=\tau)$ compared to its LRT counterpart:
quite evidently, there is a finite discontinuity in $f_k(\tau)$ which develops, for $\Delta h\to 0$, into the singular denominator 
$(\omega_0 - 2\epsilon_k^0 )^{-1}$ appearing in LRT.

\subsection{Perturbation acting on a subchain of length $l<L$} \label{subchain:sec}
%
Let us now discuss what happens if the perturbation acts only on a segment of the chain of length $l<L$,
coupling to the operator $\hat{A}=\hat{M}_l$ previously defined. 
We denote, from now on,  $\hat{m}_j=\sigma^x_j$ as the transverse magnetization at site $j$.
The LRT prediction is simple, because linearity allows us to study the response on $\hat{m}_{j'}$ to a perturbation acting on $\hat{m}_{j}$
and then appropriately summing the results. The key quantity needed is therefore $\chi_{j'j}''(\omega)$, the spectral function associated to
the retarded response function $\chi_{j'j}(t)\equiv-i\hbar^{-1}\theta(t)  \bra{\Psi_{\rm GS}} \left[\hat{m}_{j'}(t),\,\hat{m}_j\right]   \ket{\Psi_{\rm GS}}$, 
from which we can easily reconstruct the relevant $ \chi_l(t)=-i\hbar^{-1}\theta(t)\bra{\Psi_{\rm GS}}\left[\hat{M}_l(t),\,\hat{M}_l\right] \ket{\Psi_{\rm GS}}$.
Details are given in \ref{dis-loc:sec}. Note that $\chi_l$ scales as $l$ in the thermodynamic limit. 
As for the exact response of the system, we need to apply a inhomogeneous $2L \times 2L$ Bogoliubov-de Gennes theory, supplemented
by a single-particle Floquet analysis, whose technical details can be found in \ref{Bogoliubov:sec}.

\begin{figure}
\begin{center}
   \includegraphics[width=9cm]{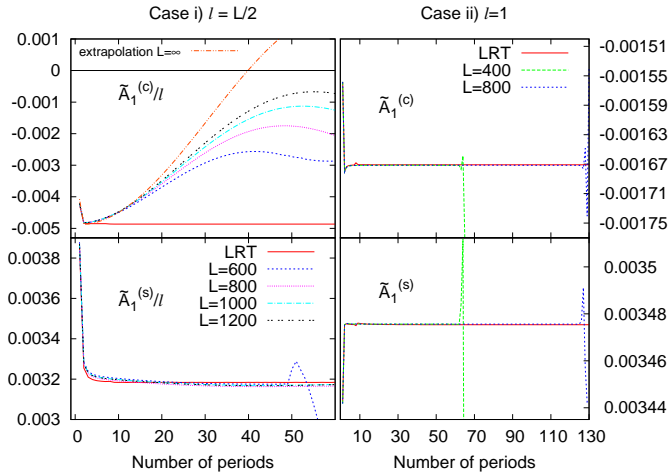}
\end{center}
\caption{(Upper panels) $\widetilde{A}_1^{(c)}(n)$, the $\cos{(\omega_0t)}$-component  of $\mean{M_l}_t$ in the $n$th-period, rescaled by $l$,
for an extensive perturbation $l=L/2$ (left, case (i)) and a local one $l=1$ (right, case (ii)). 
(Lower panels) Same as above, but for $\widetilde{A}_1^{(s)}(n)$, the $\sin{(\omega_0t)}$-component  of $\mean{M_l}_t$ in the $n$th-period.
LRT predicts $\widetilde{A}_1^{(s)}(n)=-(\Delta h/2)\chi'_l(\omega_0)$, in excellent agreement with the exact results before the revivals at $t^*$ 
for both $l=L/2$ and $l=1$.
$\widetilde{A}_1^{(c)}(n)$ quickly deviates from the LRT value $(\Delta h/2)\chi''_l(\omega_0)$ and approaches $0$ (with oscillations) for $l=L/2$, 
while it agrees with the LRT value for $l=1$. 
Here $\omega_0=1$ and $\Delta h=10^{-2}$.
}
\label{plot_comp:fig}
\end{figure}

Once again, we denote by $\widetilde{A}^{(c)}_1(n)$ the coefficient of the $\cos{(\omega_0t)}$ component of $\mean{M_l}_t$
evaluated during the $n$-th period, and by $\widetilde{A}^{(s)}_1(n)$ its $\sin{(\omega_0t)}$ component.
As discussed in Section~\ref{energy:sec}, the average energy absorption rate over the $n$-th period is given by
$\mathcal{W}_n =-(\Delta h/2) \omega_0\widetilde{A}_1^{(c)}(n)$.
Fig.~\ref{plot_comp:fig} shows the results obtained, when $\omega_0=1$ and $\Delta h=10^{-2}$, for 
an extensive perturbation with $l=L/2$ (left panels) and a local perturbation with $l=1$ (right panels).
Here the LRT results are compared with the exact ones, obtained by solving numerically the $2L\times 2L$ system of
Bogoliubov-de Gennes equations, as detailed in~\ref{Bogoliubov:sec}.
In all cases, we have studied several values of $L$ to extract the thermodynamic limit behaviour, which is as usual plagued by finite-size revival
occurring at times $t^*=2\pi n^*/\omega_0=(L-l)/v$ where $v=1$ is the group velocity of the excitations at the critical point. 
%
%
The results for $\widetilde{A}^{(s)}_1(n)$ are always in agreement with LRT, which predicts a sine-component rapidly approaching $-(\Delta h/2)\chi'_l(\omega_0)$.
The results for the cosine-component $\widetilde{A}^{(c)}_1(n)$, responsible for the energy absorption, are perfectly reproduced
by LRT, $(\Delta h/2)\chi''_l(\omega_0)$, only when the perturbation is local; on the contrary, when the perturbation is extensive, $l=L/2$, $\widetilde{A}^{(c)}_1(n)$
quickly drops to small values which likely decrease (with oscillations) towards $0$, exactly as in the $l=L$ uniform case.

A few comments regarding energy absorption are in order. LRT being obeyed at all times, when $L\to \infty$, for a local perturbation
is in some way related to the fact that the average absorption rate $-\frac{1}{8}(\Delta h)^2\omega_0\chi''_l(\omega_0)$ is a quantity of order $1$ 
which does not change the energy-per-site $(E_0(t)-E_0(0))/L$ in the thermodynamic limit. 
When $l=L/2$, on the contrary, the energy absorption predicted by LRT quickly saturates, even though half of the system might act as a ``reservoir'' for
the perturbed section. 
Finite size effects and time-revivals elucidate the mechanism behind energy absorption and LRT-failure in a physically quite transparent way. 
First of all, let us discuss the well known mechanism behind revivals in $\mean{M_l}_t$. 
The perturbation acting on the $l$-subchain generates excitations propagating along the chain at a maximum velocity $v=1$.
If the chain is infinite, the excitations will never come back;  if  $L$ is finite, due to the periodic boundary conditions, the excitations 
will return to the $l$-subchain after a time $t^*=(L-l)/v$, producing a deviation of $\mean{M_l}_t/l$ from its $L=\infty$ value. 
For $l$ finite and $L\to \infty$, the excitations would go on forever, propagating away from the perturbed sector of the chain taking away with them their initial energy~\cite{Giuliani-Vignale:book}. 
Therefore, the finite amount of energy delivered to the system in each period spreads over an infinite space: The energy per site deviates always infinitesimally from its initial value, and LRT is consequently obeyed. 
We might summarize this discussion by saying that, when the perturbation is local, it is true the common wisdom according to which an extended system acts as ``its own heat bath'' \cite{Giuliani-Vignale:book};
thus the $\eta\to 0$ factors appearing in the LRT functions are justified.

When the perturbation is extensive, $l\sim L$, the situation is very different. 
If we send $L\to \infty$ with $l/L$ constant, the ``reservoir'' $L-l$ has an infinite space over which the excitations can propagate away. 
On the other hand, as clearly indicated by the results in Fig.~\eqref{plot_comp:fig}, LRT holds only for a finite number of periods which stays finite as $L\to \infty$, and is
obviously much smaller than $n^*=t^*/\tau=(L-l)/(v\tau)$.
Evidently, the number of excitations generated by the driving in each period and the ``reservoir''-space in which they can propagate scale both with $L$:
the ``reservoir'', therefore, steadily increases its energy-per-site and the perturbation to the density matrix of the system will cease to be small: 
hence the failure of LRT, at least as far as $\chi''_l$ is concerned. 
Surprisingly, such a failure of LRT is accompanied by an excellent agreement of the $\chi'_l$-response. 
We have evidence that essentially the same picture holds for all cases with $l/L$ finite. 

One final remark concerning the local perturbation case is in order. 
Assume, for definiteness, that we perturb the system on a single site,  $\hat{A}=\hat{m}_1$, and calculate the corresponding $\mean{A}_t$.
The numerical results shown above, see Fig.~\ref{plot_comp:fig}, suggest that LRT is correct (in the limit of weak driving) at all times, 
i.e.,  $\mean{A}_t$ develops, after a transient,  an out-of-phase component proportional to $\cos{(\omega_0 t)}$, 
which is periodic but leads to a steady increase of the total energy (albeit by a non-extensive quantity). 
Referring to the general discussion of Sec.~\ref{floquet:sec}, we might ask if this periodic but out-of-phase component originates from diagonal or 
off-diagonal terms in the Floquet expansion. 
Remarkably, by exploiting the Heisenberg representation and the Bogoliubov-de Gennes equations,  
and performing a single-particle Floquet analysis of the latter, see \ref{Bogoliubov:sec}, we have a numerical way of 
extracting $\mean{A}_t^{\rm diag}$, $\mean{A}_t^{\rm off-diag}$ and its spectral density $F_t(\omega)$, 
see Eqs.~\ref{aver-per}-\ref{aver-fluc}-\ref{outofdiag},
which in principle involve many-body matrix elements and Floquet quasienergies. 
Our numerical analysis suggests that, for every finite size $L$, there are two-fold quasi-degeneracies of single-particle Floquet quasienergies $\mu_{\alpha}$,
(i.e., for every $\alpha$ there is a $\bar{\alpha}\neq \alpha$ such that $\mu_{\bar \alpha} \sim \mu_{\alpha}$) which likely become strict degeneracies for $L\to \infty$,  
and which appear to be a possible source of a singularity in the spectral function $F_t(\omega\to 0)$,
thus violating the hypothesis of the Riemann-Lebesgue lemma and giving rise to a persisting out-of-phase contribution.

Summarizing, for a localized perturbation and in the long-time limit, the terms in $m_1(t)$ which are diagonal in the Floquet basis contribute 
only to the in-phase response.
Quasi-degenerate off-diagonal terms give a further contribution to the in-phase response, as well as the entire out-of-phase response. 
These off-diagonal quasi-degenerate contributions to $m_1(t)$ ultimately lead to a periodic response, matching LRT, up to a time $\tilde{t}$ of 
the same order of the inverse gap among the quasi-degenerate Floquet levels, hence for longer and longer $\tilde{t}$ as $L\to \infty$.
%
This fact mirrors the physical picture that the space in which we can accomodate excitations grows to infinity in this limit.


\section{Discussion and conclusions} \label{conclusions:sec}
%
The results discussed above have been explicitly demonstrated, so far, just for an Ising chain with a periodically modulated transverse field
around the critical point. It is natural to ask how robust they are in more general circumstances.

 
The system we explicitly discuss is essentially a free-fermion (BCS) problem. Would interactions between fermions modify this result? 
Although we have no mathematical proof for this, we believe that this is not the case. 
A circumstantial evidence for this claim comes from the numerical results of Ref.~\cite{Kollath_PRA06} where a Hubbard chain 
with a hopping which is periodically modulated in time 
--- mimicking fermionic cold atoms experiments --- is studied using t-DMRG \cite{Daley_JSTAT04,White_PRL04}: 
the energy absorbed by the system shows clear signs of a saturation similar to that of our Figure \ref{workers:fig}.
Admittedly, a fermionic one-dimensional Hubbard model is still integrable (by Bethe-{\em Ansatz}) in equilibrium, but we believe that
integrability is not a crucial issue in the present context: what we believe crucial (see discussion in Section \ref{energy:sec})
is that there is a maximum energy-per-site $\epsilon_{\rm max}$  that the system can have, so that $\langle \Psi(t) | H_0 |\Psi(t)\rangle < L \epsilon_{\rm max}$ 
at all times, whereas LRT predicts, when $\chi''\neq 0$, a steady increase of energy for large $t$.
In view of the energy considerations of Section \ref{energy:sec}, we believe that our results apply, both for extensive and local perturbations, 
whenever the energy-per-site spectrum is bounded; this condition is verified for all the rigid lattice systems. 

A word of caution applies to systems (for example a bosonic Hubbard model) that do not have a bound on the maximum energy-per-site.
An obvious counter-example to our discussion is that of a system of driven harmonic oscillators 
(masses interacting with nearest-neighbor springs and subject, for instance, to a localized periodic perturbation
$E(t) x_1$)\footnote{Here $E(t)$ mimicks an electric field acting locally on a single particle, assumed to posses a dipole moment.}.
The linearity of the problem, indeed, makes LRT exact at all times, implying that the system will steadily increase its energy in time 
when the frequency $\omega_0$ of the driving falls inside the natural spectral range of the problem.  
At the linear level, obviously, Ehrenfest theorem guarantees that quantum and classical physics results coincide. 
When non-linearities are included, for instance adding cubic nearest-neighbor interactions, as in the Fermi-Pasta-Ulam problem \cite{art:Fermi-Pasta-Ulam},
interesting questions emerge concerning classical \cite{Castiglione:book} versus quantum non-equilibrium physics, and deviations from LRT. 
Although we do not have a full picture of this problem, simulations we have conducted on the classical Fermi-Pasta-Ulam chain 
with a localized periodic perturbation suggest that, when the non-linearity is strong enough, there are marked deviations from LRT but in
such a way that the energy increases in time in a ``stronger-than-linear'' way, quite differently from the saturation effects previously
described for quantum systems on a lattice. 
Regarding classical versus quantum physics in the phenomena of interest, we stress that both the bounded energy-per-particle spectrum 
as well as the role of off-diagonal matrix elements with the accompanying dephasing, are intrinsically quantum ingredients:
the effects described, therefore, might not survive in the classical regime. 
Equally deserving further study are quantum problems on the continuum --- where no single-band cut-off, 
typical of lattice problems, applies ---,  as well as the case of lattice systems in the presence of phononic modes.
In the first case the answer is not obvious: for instance electrons moving in a continuum crystalline potential have a band energy spectrum 
without an upper bound; though in some cases~\cite{Landauer_PRB86,Gefen_PRL87} quantum coherence effects still forbid energy absorption beyond a certain limit.
We observe also that there is a similarity of our results with dynamical localisation \cite{Stoeckmann:book} (quantum coherence and saturation), 
but in our case a thermodynamic limit is essential, while dynamical localisation generally applies to systems whose unperturbed spectrum 
is characterized by a discrete level spacing.

%
%
%
Finally, let us stress once more the striking difference between a driving which acts {\em locally}, where LRT appears
to apply at all times, and a driving involving an extensive perturbation. Evidently, no perturbation can be considered to be ``small'' at all times
unless the system can act as a ``its own bath'', which implies that the perturbation should not modify in any essential way 
the energy-per-site: if there is an infinite space in which the finite number of excitations generated by the driving
in each period can propagate, the excitation energy per site will be always infinitesimal.
On the contrary, when the perturbation is extensive, the energy pumped into the system, if no mechanism for dissipation is provided, will lead
to a failure of LRT after a certain finite time: surprisingly enough there are quantities, like the in-phase response proportional to $\chi'(\omega_0)$
which are well described by LRT at all times.
A non-trivial case might be constituted by systems with localized states, where the excitations generated by a local perturbation,
due to the absence of diffusion implied by the localization, cannot propagate away from the perturbed region: the local energy growth
might then drive the system away from LRT.

Are the results we have discussed of any relevance to experiments? 
Obviously, no physical system is perfectly closed: coupling to an environment always leads to decoherence, take for instance the uncontrolled
interactions with the electromagnetic field of cold atoms in optical lattices, or the coupling of electronic degrees of freedom in a solid to the phononic modes
of the lattice. 
Nevertheless the evolution can be considered unitary until correlations with the environment set up: this happens 
after a time scale which modern experimental techniques can resolve. For instance, in experiments with cold atoms in optical lattices
coherence times have been attained of $\sim 1$ ms~\cite{Sias_PRL08,Lignier_PRL07}; we think that taking a trapped systems of about $10^4$ atoms
 (for which we can reasonably talk about a ``thermodynamic limit'') a periodic perturbation can be realised and in principle, with an appropriate 
choice of $\omega_0$, a regime can be reached in which LRT is expected to hold and where the afore-discussed effects can be checked.
In the solid state, the dynamics of electrons stays coherent for much shorter time-scales, $\sim 1$ ps; nevertheless, even such extremely short time-scales
are in principle within the experimental reach of modern ultrafast pump-and-probe spectroscopic techniques~\cite{Enciclopedia:book,Shah:book,Glezer:phdthesis,Nasu:book}.

\appendix
\section{} \label{singularity-LR:app}
%
In this appendix we examine the singularities of the LRT susceptibility in the light of the standard textbook approach, which 
includes an adiabatic switching-on factor for $t\in (-\infty,0]$. Consider a periodic perturbing field which is turned on at $-\infty$
as:
\begin{equation}
v(t) = v_{\rm switch}(t) + v_{\rm per}(t) = v_0 \sin(\omega_0 t) \left[ \nep^{\eta t} \theta(-t) + \theta(t) \right] \;,
\end{equation}
where $\eta\to 0$ at the end of the calculation, and define $\delta \mean{A}_t \equiv \mean{A}_t - \mean{A}_{\rm eq}$.
Since we will consider only the linear terms in $v$, we can calculate the two terms separately and add the results.
The switching-on part $v_{\rm switch}(t) = v_0  \theta(-t) \nep^{\eta t} \sin{(\omega_0 t)}$ leads, for $t\ge 0$ and $\eta\to 0$, to:
\begin{equation}  \label{meaner-adiab}
  \delta\mean{A}_t^{\rm switch} = 
  v_0 \dashint_{-\infty}^{+\infty} \!\! \frac{\ud\omega}{2\pi i} \;   
                          \left( \frac{\chi''(\omega)}{\omega + \omega_0} - \frac {\chi''(\omega)}{\omega-\omega_0} \right) \nep^{-i\omega t}  
  - v_0 \chi''(\omega_0) \cos{ (\omega_0 t)} \;,
\end{equation}
where we made use of the standard approach for  dealing with poles in terms of Cauchy principal-value integrals and Dirac's deltas:
\begin{equation}
\lim_{\eta\to 0} \int_{-\infty}^{+\infty} \! \! \ud\omega \, \frac{f(\omega)}{\omega-\omega_0+i\eta} = 
\dashint_{-\infty}^{+\infty} \! \! \ud\omega \, \frac{f(\omega)}{\omega-\omega_0} - i\pi f(\omega_0) \;. \nonumber
\end{equation}
It is clear that the first integral will have to cancel, for large $t$, the second term, because, physically,   
$\delta\mean{A}_t^{\rm switch}$ represents the relaxation towards equilibrium after the field was turned on 
in $(-\infty,0]$. 
Before proceeding with the (simple) mathematical justification of this statement, let us comment that the Cauchy principal value
integral appearing in Eq.~\eqref{meaner-adiab} is exactly the same, with an opposite sign, as that appearing in the
expression for $\delta\mean{A}_t^{\rm per}$ derived in Section \ref{linear-response:sec}, since
\begin{equation} \label{equality:eqn}
 \dashint_{-\infty}^{+\infty} \!\! \frac{\ud\omega}{2\pi i} \;   
                          \left( \frac{\chi''(\omega)}{\omega + \omega_0} - \frac {\chi''(\omega)}{\omega-\omega_0} \right) \nep^{-i\omega t}  
= 2 \omega_0 \dashint_{0}^{+\infty} \! \! \frac{\ud\omega}{\pi} \; \frac{\chi''(\omega)}{\omega^2-\omega_0^2} \sin{(\omega t)} \;.
\end{equation}
Therefore, if we sum the two terms we obtain the total response to $v(t)$ as:
\begin{equation}  \label{total-response:eqn}
  \delta\mean{A}_t =   v_0 \left[ \chi'(\omega_0) \sin{ (\omega_0 t)}  - \chi''(\omega_0) \cos{ (\omega_0 t)} \right] \;, 
\end{equation}
as indeed expected.

We now show that:
\begin{equation} \label{Singular-int:eqn}
v_0 \dashint_{-\infty}^{+\infty} \!\! \frac{\ud\omega}{2\pi i} \;   
                          \left( \frac{\chi''(\omega)}{\omega + \omega_0} - \frac {\chi''(\omega)}{\omega-\omega_0} \right) \nep^{-i\omega t}  
                          = v_0 \chi''(\omega_0) \cos{(\omega_0 t)} + F^{\rm relax}(\omega_0,t) \;,
\end{equation}
where $F^{\rm relax}(\omega_0,t)$ is a function which relaxes to $0$ for $t\to \infty$.
First, we see from Eq.~\eqref{chisec} that $\chi''(\omega)$ is non-vanishing only when $\omega$ matches a resonance 
frequency of the system. 
We assume we are dealing with a system whose resonance spectrum is a smooth continuum, in which case $\chi''(\omega)$ is a regular function. 
The function $\chi''(\omega)$ is odd in $\omega$, so if $\omega_0$ falls inside the resonance spectrum $\chi''(-\omega_0)=-\chi''(\omega_0)\neq 0$; if
it falls outside $\chi''(\pm\omega_0)= 0$. In both cases we can formally split the first term in the integrand (the second term can be
treated in the same way)
%
%
%
\begin{equation}
  \frac{\chi''(\omega)}{\omega + \omega_0}\nep^{-i\omega t} = \frac{\chi''(\omega)-\chi''(-\omega_0)}{\omega + \omega_0}\nep^{-i\omega t}
  +\frac{\chi''(-\omega_0)}{\omega + \omega_0} \nep^{-i\omega t} \;.
\end{equation}
The first term is always regular, even for $\omega\to-\omega_0$, and it leads to an integral that vanishes for large $t$ (Riemann-Lebesgue lemma). 
Whenever $\chi''(\pm\omega_0)\neq 0$, the second term is singular in $-\omega_0$ and contributes to the integral with the piece
%
\begin{equation} \label{integro}
  \chi''(-\omega_0) \dashint_{-\infty}^{+\infty} \!\! \frac{\ud\omega}{2\pi i} \; \frac{\nep^{-i\omega t}}{\omega + \omega_0} \;.
\end{equation}
Because of the singularity, this integral does not vanish in the long-time limit, as we are going to show evaluating it with the usual complex plane techniques. 
Assuming $t>0$, we can close the integration contour, both at infinity and around the singularity, in the lower half complex semi-plane, as shown in Figure \ref{path:fig}. 
%
%
Using standard techniques, one concludes that the principal-value integral we need is given by (minus) the contribution around the singularity ($-i\pi\nep^{i\omega_0 t}/(2\pi i)$),
hence:
%
\begin{figure}
\begin{center}
   \includegraphics[width=8cm]{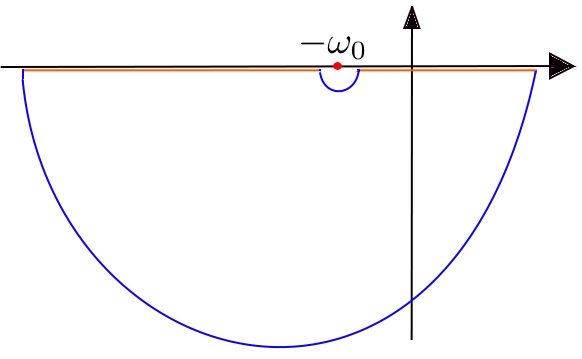}
\end{center}
\caption{The integration contour used to evaluate the principal value integral in Eq.~\eqref{integro}.}
\label{path:fig}
\end{figure}
%
\begin{equation}
  \chi''(-\omega_0) \dashint_{-\infty}^{+\infty} \!\! \frac{\ud\omega}{2\pi i} \; \frac{\nep^{-i\omega t}}{\omega + \omega_0} =
   - \frac{\chi''\left(-\omega_0\right)}{2}\nep^{i\omega_0 t}\,.
\end{equation}
By repeating this argument for the term with the pole at $\omega_0$ and exploiting the fact that $\chi''(\omega)$ is odd in $\omega$,
one finally arrives at Eq.~\eqref{Singular-int:eqn}, where $F^{\rm relax}$ is explicitly given by:
\begin{equation}  \label{meaner-adiab-2}
  F^{\rm relax}(\omega_0,t) = 
  v_0 \int_{-\infty}^\infty \! \! \frac{\ud\omega}{\pi} \; \frac{[\chi''(\omega)-\chi''(\omega_0)]}{\omega-\omega_0} \sin{(\omega t)}  \;.
\end{equation}
Notice, finally, that $F^{\rm relax}(\omega_0,t)= -F^{\rm trans}(\omega_0,t)$, where $F^{\rm trans}(\omega_0,t)$ is the transient term appearing in
Eqs.~\eqref{LR-transient:eqn}-\eqref{LR-transient-2:eqn}, and $\delta\mean{A}_t^{\rm switch} =  F^{\rm relax}(\omega_0,t)$.

\section{} \label{trasmag:app}
%
In this appendix we evaluate the zero-temperature transverse magnetisation density for a Ising chain within linear response theory.
The response function we need to calculate is (with $\hbar=1$):
\begin{equation} \label{response}
	\chi(t) = - i \theta(t) \,\bra{\Psi_{\rm GS}}\left[\hat{m}(t),\hat{M}\right]\ket{\Psi_{\rm GS}}   
		 = -i \theta(t) \frac{1}{L} \sum_{k>0}^{\textrm{ABC}} \bra{\psi_0^k} \left[\hat{m}_k(t), \hat{m}_k\right] \ket{\psi_0^k} \;,
\end{equation}
where $\hat{m}_k(t) = 2 \left( c_{-k}(t) c_{-k}^\dagger(t) - c_k^\dagger(t) c_k(t) \right)$ is a Heisenberg's operator evolving with
$\hat{H}_0$, see \ref{H0:eqn},  $\hat{m}_k=\hat{m}_k(0)$, and we have exploited the fact that the different $k$-subspaces are perfectly decoupled. 
The ground state $\ket{\Psi_{\rm GS}}$ of $\hat{H}_0$ is given by Eq.~\eqref{ground}
in which $u_k^{0}=\cos(\theta_k/2)$ and $v_k^{0}=i\sin(\theta_k/2)$ with $\tan{\theta_k} = (\sin{k})/(1-\cos{k})$.
To find $\hat{m}_k(t)$ we need $c_k(t)$, which obeys a Heisenberg's equation of motion with Hamiltonian $\hat{H}_0$ and initial value $c_k(0)=c_k$.
It is simple to derive that $c_k(t) = p_k(t) c_k + q_k(t) c_{-k}^\dagger$ with 
%
%
$p_k(t) =  \cos(\epsilon_k^{0} t) - i\cos(\theta_k) \sin(\epsilon_k^{0} t)$, $q_k(t) = -\sin(\theta_k) \sin(\epsilon_k^{0} t)$, 
and $\epsilon_k^0=2\sin(k/2)$. 
With these ingredients it is a matter of simple algebra to derive the following expression for $\chi(t)$:
\begin{equation}    \label{itski}
	\chi(t) = - \theta(t) \frac{8}{L} \sum_{k>0}^{\textrm{ABC}} \cos^2\left(\frac{k}{2}\right) \sin(2\epsilon_k^{0} t) \;,
\end{equation}
which in turn immediately gives, by Fourier transforming:
\begin{equation}    \label{chiz-Ising:eqn}
	\chi(z) = - \frac{4}{L} \sum_{k>0}^{\textrm{ABC}} \cos^2\left(\frac{k}{2}\right) 
	\left[   \frac{1}{2\epsilon_k^0 -z} +  \frac{1}{2\epsilon_k^0 +z}  \right]  \;.
\end{equation}
The spectral function $\chi''(\omega)$ can be directly extracted from this expression:
\begin{equation}    \label{imchi-Ising:eqn}
	\chi''(\omega>0) = - \frac{4\pi}{L} \sum_{k>0}^{\textrm{ABC}} \cos^2\left(\frac{k}{2}\right) 
	\delta \left( \omega - 2\epsilon_k^0 \right) \stackrel{\scriptscriptstyle L\to \infty} {\longrightarrow}
	-  \theta(4-\omega) \, \sqrt{1-\left(\frac{\omega}{4}\right)^2} \;,
\end{equation}
where we have taken the thermodynamic limit  ($\frac{1}{L}\sum_{k>0}^{\rm ABC} \to  \int_0^\pi \frac{dk}{2\pi}$) which
transforms the discrete sum of Dirac's delta functions into a smooth function.

It is worth mentioning the finite-size LRT expression for $\delta \mean{m}_t$, immediately obtained from Eq.~\eqref{itski}:
\begin{equation} \label{magliner}
  \delta \mean{m}_t =-\Delta h \frac{4}{L} \sum_{k>0}^{\textrm{ABC}} \cos^2\left(\frac{k}{2}\right)
  \frac{2\epsilon_k^{0} \sin\left(\omega_0 t\right) -\omega_0 \sin\left(2\epsilon_k^{0}t\right)}{\omega_0^2-\left(2\epsilon_k^{0}\right)^2} \;.
\end{equation}
At finite size, there are discrete isolated resonances occurring when $\omega_0$ coincides with one of the excitation 
frequencies of the unperturbed system: $\omega_0=2\epsilon_{\bar k}^0$. 
Such a resonance gives rise to a quite unphysical prediction of LRT: there is a contribution to $\delta \mean{m}_t$ 
originating from the $\bar k$-term in the sum over $k$ which can be shown (using  {\it de l'H\^opital} theorem) to
grow without bounds in time as $-2(\Delta h) L^{-1} \cos^2({\bar k}/2) \; t \cos(\omega_0 t)$. 
Notice that this divergent contribution carries a $1/L$ factor. The amusing thing coming out of the thermodynamic limit
is that such isolated resonances are, in some sense, transformed into ``principal value singularities'' which do
not give rise to any divergence in $\delta \mean{m}_t$, although they are, in the end, responsible for the out-of-phase
contribution to $\delta \mean{m}_t$, proportional to $\chi''(\omega_0)$, which we have discussed in the text.
%
%
\section{} \label{dis-loc:sec}

In this section we discuss the local susceptibility $\chi_{j0}$. 
The local magnetisation operators are defined as $\hat{m}_j\equiv\sigma_j^x$, and the response function we are interested in can be
written as
\begin{equation}
  \chi_{j0}(t) \equiv -\frac{i}{\hbar}\theta(t) \bra{\Psi_{\rm GS}} \left[\hat{m}_j(t),\,\hat{m}_0\right] \ket{\Psi_{\rm GS}} \,.
\end{equation}
As mentioned in Section~\ref{linear-response:sec}, the crucial information is contained in $\chi_{j0}''(\omega)$ 
which reads:
\begin{equation} \label{chi_io?}
  \chi_{j0}''(\omega) = -\frac{\pi}{\hbar} \sum_{n\neq0}\left[ ({m}_j)_{n0}^*({m}_0)_{n0} \,\delta\left(\omega-\omega_{n0}\right)
                                -({m}_j)_{n0}({m}_0)_{n0}^* \,\delta\left(\omega-\omega_{0n}\right) \right]\,,
\end{equation}
where the sum extends over the eigenstates (0 labels the ground state); the matrix elements $(m_j)_{mn}$ and the frequencies $\omega_{mn}$
are defined as in Eq.~\eqref{chicchiricchi}. As $\chi_{j0}''(\omega)$ is odd in $\omega$, we need to consider only $\omega\ge 0$.
Using the Jordan-Wigner transformation we can write
\begin{equation} \label{mjel}
  ({m}_j)_{n0} = \bra{n}\hat{m}_j \ket{\Psi_{\rm GS}}  = -\frac{2}{L}\sum_{k,\,k'} \bra{n} c_k^\dagger c_{k'} \ket{\Psi_{\rm GS}} \nep^{i(k'-k)j}
\end{equation}
where the fermionic operators $c_k$ have been defined in Section~\ref{Ising:sec}. The operators $\gamma_k$ diagonalising the quadratic Hamiltonian
Eq.~\eqref{H0:eqn} can be obtained from the $c_k$ with a Bogoliubov transformation 
$c_k=u_k^0\gamma_k+v_k^0\gamma_{-k}^\dagger$, $ c_{-k}^\dagger=-{v_k^0}^*\gamma_k+u_k^0\gamma_{-k}^\dagger$.
%
%
If we substitute in Eq.~\ref{mjel} we see that the only non-vanishing matrix element is among the ground state and excited states whose form
is $\gamma_{\widetilde{k}'}^\dagger\gamma_{\widetilde{k}}^\dagger \ket{\Psi_{\rm GS}}$. Applying Wick's theorem we can write
\begin{equation}
  ({m}_j)_{n0} = -\frac{2}{L} \sum_{k,\,k'} \bra{n} c_k^\dagger c_{k'} \ket{\Psi_{\rm GS}}  \nep^{i(k'-k)j} = -\frac{2}{L}\left(-u_{\widetilde{k}'}^0v_{\widetilde{k}}^0+
  u_{\widetilde{k}}^0v_{\widetilde{k}'}^0\right)\nep^{-i(\widetilde{k}+\widetilde{k}')j} \,,
\end{equation}
where we have exploited that $v_{-k}^0=-v_{k}^0$ and $u_{-k}^0=u_{k}^0$. Substituting this expression in Eq.~\eqref{chi_io?}, and using that for the
relevant excited states $\omega_{n0}=\epsilon_{\widetilde{k}}+\epsilon_{\widetilde{k}'}$ we can write
\begin{equation} \label{chion}
  \chi_{j0}''(\omega\ge 0) = -\frac{\pi}{\hbar}\frac{4}{L^2}\sum_{\widetilde{k}>\widetilde{k}'}\left|u_{\widetilde{k}}^0v_{\widetilde{k}'}^0-
u_{\widetilde{k}'}^0v_{\widetilde{k}}^0\right|^2\nep^{-i(\widetilde{k}+\widetilde{k}')j}
\delta(\omega-\epsilon_{\widetilde{k}}^0-\epsilon_{\widetilde{k}'}^0) \;,
\end{equation}
where the condition $\widetilde{k}>\widetilde{k}'$ has been enforced to avoid double counting of the excited states $\ket{n}$.
The object inside the sum is symmetric upon exchange of $\widetilde{k}$ and $\widetilde{k}'$. 
%
%
Using this, restricting the sum to the positive $\widetilde{k}$ and $\widetilde{k}'$ and going to the thermodynamic limit we get:
%
%
%
\begin{eqnarray}
\hspace{-20mm} \chi_{j0}''(\omega\ge 0) &=& -\frac{4}{\pi\hbar} \int_0^\pi\ud {k} \int_0^\pi \ud {k}'
\left\{ \left| u_{k}^0 v_{k'}^0 \right|^2 \cos{(kj)} \cos{(k'j)} \right. \nonumber \\
&& \left. \hspace{15mm} - u_{k'}^0 v_{k}^0 u_{k}^0 v_{k'}^0 \sin{(kj)} \sin{(k'j)} \right\}
\delta(\omega-\epsilon_{k}^0-\epsilon_{k'}^0) \,. \nonumber
\end{eqnarray}
Using the expressions for $u_k^0$ and $v_k^0$ in Section~\ref{Ising:sec} and changing variable to $\epsilon=2\sin(k/2)$,  
we can rewrite this as:
\begin{eqnarray}
\hspace{-20mm}  \chi_{j0}''(\omega\ge 0) &=& -\frac{1}{\pi\hbar} \int_{\max(0,\omega-2)}^{\min(\omega,2)} \ud\epsilon \bigg[\sqrt{\frac{(2-\epsilon)(2+\epsilon-\omega)}
{(2+\epsilon)(2+\omega-\epsilon)}} \cos(k_\epsilon j)\cos(k_{\omega-\epsilon}j) \nonumber \\
&& \hspace{35mm} + \sin(k_\epsilon j) \sin(k_{\omega-\epsilon}j ) \bigg]\,,
\end{eqnarray}
where we have defined the function $k_\epsilon\equiv 2\arcsin(\epsilon/2)$. 
%

The linear response function needed in the text is obtained from $\chi_{j0}$ via the expression:
\begin{equation}
  \chi_l(t) = -\frac{i}{\hbar} \theta(t) \bra{\Psi_{\rm GS}} \left[ \hat{M}_l(t),\,\hat{M}_l\right] \ket{\Psi_{\rm GS}}  = l\sum_{j=-l+1}^{l-1}\chi_{j0}(t) \,.
\end{equation}
Observe that cancellations in the sum over $j$, due to the highly oscillating contributions $\chi_{j0}(t)$, make $\chi_l$ proportional to $l$ 
rather than to $l^2$.

\section{} \label{Bogoliubov:sec}
%
In this Appendix we briefly describe the quantum dynamics of inhomogenous Ising/XY chains \cite{Caneva_PRB07}. 
Generically, if $c_j$ denote the $L$ fermionic operators originating from the Jordan-Wigner transformation of spin operators,
we can write the Hamiltonian in Eq.~\eqref{h11} as a quadratic fermionic form
\begin{equation} \label{hamor}
 \hat{H}(t) = 
\hat{\mathbf{\Psi}}^{\dagger} \cdot {\mathbb H}(t) \cdot \hat{\mathbf{\Psi}}  
  = \left( \begin{array}{cc}  {\bf c}^{\dagger} & {\bf c} \end{array} \right)
  \left( \begin{array}{rr} {\bf A}(t) & {\bf B}(t) \\
                                        -{\bf B}(t) & -{\bf A}(t) \end{array} \right)
                                        \left( \begin{array}{l}  {\bf c} \\ {\bf c}^\dagger \end{array} \right)
                                        \;,
\end{equation}
where $\hat{\mathbf{\Psi}}$ are $2L$-components (Nambu) fermionic operators defined as 
$\Psi_j = c_j$  (for $1\le j\le L$) and $\Psi_{L+j} = c_j^\dagger$,  
and  ${\mathbb H}$ is a $2L\times 2L$ Hermitean matrix having the explicit form shown on the right-hand side, 
with $\bf A$ an $L\times L$ real symmetric matrix, $\bf B$ an $L\times L$ real anti-symmetric matrix. 
Such a form of ${\mathbb H}$ implies a particle-hole symmetry: if $({\bf u}_{\alpha}, {\bf v}_{\alpha})^T$ is an instantaneous eigenvector 
of ${\mathbb H}$ with eigenvalue $\epsilon_{\alpha}\ge 0$, then $(-{\bf v}_{\alpha}^*, {\bf u}_{\alpha}^*)^T$ is an eigenvector with 
eigenvalue $-\epsilon_{\alpha}\le 0$.

Let us now focus on a given time, $t=0$, or alternatively suppose that the Hamiltonian is time-independent. 
Then, we can apply a unitary Bogoliubov transformation  
\begin{equation} \label{trasgro}
  \hat{\mathbf{\Psi}} = \left( \begin{array}{l}  {\bf c} \\ {\bf c}^\dagger \end{array} \right) = 
  {\mathbb U}_0 \cdot \left( \begin{array}{l} \bgamma \\ \bgamma^\dagger \end{array} \right) =
  \left(\begin{array}{rr} {\bf U}_0 & -{\bf V}^*_0 \\
                                       {\bf V}_0 & {\bf U}^*_0 \end{array} \right) \cdot  
                                       \left( \begin{array}{l} \bgamma \\ \bgamma^\dagger \end{array} \right) \;,
\end{equation}
where ${\bf U}_0$ and ${\bf V}_0$ are $L\times L$ matrices collecting all the eigenvectors of $\mathbb H$, by column, turning the Hamiltonian 
in Eq.~\eqref{hamor} in the diagonal form
\begin{equation}
  \hat{H} = \sum_{\alpha=1}^L \epsilon_\alpha \left(\gamma_{\alpha}^\dagger \gamma_{\alpha} - \gamma_{\alpha}\gamma_{\alpha}^\dagger\right)\,,
\end{equation}
where the $\gamma_{\alpha}$ are new quasiparticle Fermionic operators. 
%
%
The ground state $\ket{{\rm GS}}$ has energy $E_{\rm GS}=-\sum_\alpha \epsilon_\alpha$ and is the vacuum of the $\gamma_{\alpha}$ for all values of 
$\alpha$: $\bra{{\rm GS}}\gamma_{\alpha}^\dagger \gamma_{\alpha}\ket{{\rm GS}}=0$. 
\footnote{We notice that it would be easy to implement a coherent evolution of a system initially in thermal equilibrium at temperature 
$T=1/(k_B\beta)$, by imposing at time $t=0$ that $\mean{\gamma_{\alpha}^\dagger \gamma_{\alpha}}_0=\frac{1}{\nep^{\beta \epsilon_\alpha}+1}$ 
and going on with the following analysis.}

To discuss the quantum dynamics when ${\hat H}(t)$ depends on time, one starts by writing the Heisenberg's equations of
motion for the $\hat{\mathbf{\Psi}}$, which turn out to be {\em linear}, due to the quadratic nature of $\hat{H}(t)$.
A simple calculation shows that:
\begin{equation}
i\hbar \frac{d}{dt} \hat{\mathbf{\Psi}}_{H}(t) = 2 {\mathbb H}(t) \cdot \hat{\mathbf{\Psi}}_{H}(t) \;,
\end{equation}
the factor $2$ on the right-hand side originating from the off-diagonal contributions due to $\{\Psi_j,\Psi_{L+j}\}=1$.
These Heisenberg's equations should be solved with the initial condition that, at time $t=0$, is
\begin{equation} \label{trasgro_0}
  \hat{\mathbf{\Psi}}_H(t=0) = \hat{\mathbf{\Psi}} = 
{\mathbb U}_0 \cdot  \left( \begin{array}{l} \bgamma \\ \bgamma^\dagger \end{array} \right) \;.
\end{equation}
A solution is evidently given by 
\begin{equation} \label{Psi-Heis:eqn}
  \hat{\mathbf{\Psi}}_H(t) = 
{\mathbb U}(t) \cdot  \left( \begin{array}{l} \bgamma \\ \bgamma^\dagger \end{array} \right)
\end{equation}
with the same $\bgamma$ used to diagonalize the initial $t=0$ problem, as long as the time-dependent coefficients
${\mathbb U}(t)$ satisfy the ordinary linear Bogoliubov-de Gennes time-dependent equations:
\begin{equation} \label{bog}
i\hbar \frac{d}{dt} {\mathbb U}(t) 
= 2{\mathbb H} (t) \cdot 
{\mathbb U}(t) 
\end{equation}
with initial conditions ${\mathbb U}(t=0)={\mathbb U}_0$. 
It is easy to verify that the time-dependent Bogoliubov-de Gennes form implies that the operators $\gamma_{\alpha}(t)$ in 
the Schr\"odinger picture are time-dependent and annihilate the time-dependent state $|\psi(t)\rangle$.
Notice that ${\mathbb U}(t)$ looks like the unitary evolution operator of a $2L$-dimensional problem with Hamiltonian $2{\mathbb H}(t)$.
This implies that one can use a Floquet analysis to get ${\mathbb U}(t)$ whenever ${\mathbb H}(t)$ is time-periodic. 
This trick provides us with {\em single-particle} Floquet modes and quasi-energies in terms of which we can reconstruct,
through the Heisenberg picture prescription, the expectation value of an operator $\langle \psi(t) | \hat{O} |\psi(t)\rangle$:
it is enough to express $\hat{O}$ in terms of the fermions $\Psi_j$, and then use the Heisenberg picture and the (numerical) solution of
the Bogoliubov-de Gennes equations.
%
%
For instance, for the transverse magnetization 
$\hat{m}=\frac{1}{L}\sum_{j=1}^{L}\sigma_j^x=\frac{1}{L}\sum_{j=1}^{L}\left(1-2c_j^\dagger c_j\right)$ we immediately get:
%
%
%
\begin{equation} \label{magvalue}
   m(t) = \langle \psi(t) | \hat{m} | \psi(t) \rangle = 
   1-\frac{1}{L}\sum_{j,\,\alpha=1}^L \big( \left| U_{j\,\alpha}(t)\right|^2 \mean{ \gamma_{\alpha}^\dagger \gamma_{\alpha}}_0
        + \left| V_{j\,\alpha}(t) \right|^2 \mean{\gamma_{\alpha} \gamma_{\alpha}^\dagger}_0 \big) \;,
\end{equation}
where $U_{j\alpha}(t)=[{\mathbb U}(t)]_{j,\alpha}$ and $V_{j\alpha}(t)=[{\mathbb U}(t)]_{L+j,\alpha}$.
By expanding the $U_{j\,\alpha}(t)$, $V_{j\,\alpha}(t)$ in the corresponding single-particle Floquet modes, we can easily isolate the periodic 
and the fluctuating part of $m(t)$.

Further details on the practical implementation of this procedure for the homogeneous Ising case are given in
the Supplementary Material of Ref.~\cite{Russomanno_PRL12}.
%
%
%
In the inhomogenous case, we aim to find the evolution matrix over one period $\tau$, ${\mathbb U}(\tau)$, of the $2L\times2L$ 
Bogoliubov-de Gennes equations Eq.~\eqref{bog}. 
Particle-hole symmetry 
\footnote{Notice that, due to the particle-hole form of ${\mathbb H}(t)$, it is enough to solve 
\begin{equation} \label{bog-2}
i\hbar \frac{d}{dt} \left( \begin{array}{c} {\bf U}(t)  \\ {\bf V}(t)  \end{array} \right) 
= 2 {\mathbb H} (t) \cdot \left( \begin{array}{c} {\bf U}(t)  \\ {\bf V}(t)  \end{array} \right) \;,
\end{equation}
the full ${\mathbb U}(t)$ being given by:
\[ 
{\mathbb U}(t) = \left( \begin{array}{rr} {\bf U}(t)  & -{\bf V}^*(t) \\ {\bf V}(t)  & {\bf U}^*(t) \end{array} \right) \;.
\]
}
simplifies our job allowing us to solve those equations for $L$ different initial conditions 
$\big(\underbrace{1,\dots,0}_{L}\big|\underbrace{0,\dots,0}_{L}\big)^T,\,\dots,\,\big(\underbrace{0,\dots,1}_{L}\big|\underbrace{0,\dots,0}_{L}\big)^T$.
Diagonalizing the ${\mathbb U}(\tau)$ so constructed, 
we obtain the quasi-energies as the phases of the eigenvalues. 
\footnote{For numerical reasons, it is better to diagonalize the $2L\times 2L$ Hermitean matrix
\begin{equation}
  {\mathbb A} = -i\left(\mathbf{1}-{\mathbb U}(\tau)\right) \left(\mathbf{1}+{\mathbb U}(\tau)\right)^{-1} \;.
\end{equation}
The Floquet quasi-energies are obtained from the $2L$ eigenvalues $a_\alpha$ of ${\mathbb A}$ as 
$\mu_\alpha = \frac{\omega_0}{\pi}\atan a_\alpha$. 
}
%

\ack 
We acknowledge discussions with M. Fabrizio, C. Kollath, J. Marino, G. Menegoz, P. Smacchia, E. Tosatti and S. Ziraldo.
Research was supported by MIUR, through PRIN-2010LLKJBX-001, by SNSF, 
through SINERGIA Project CRSII2 136287\ 1, by the EU-Japan Project LEMSUPER, and 
by the EU FP7 under grant agreement n. 280555.
GES dedicates this paper to the dear memory of his friend and mentor Gabriele F. Giuliani.


\section*{References}

\begin{thebibliography}{10}

\bibitem{Pines-Nozieres:book}
D.~Pines and P.~{Nozi\`eres}.
\newblock {\em The theory of quantum liquids}.
\newblock W.A. Benjamin, Inc., 1966.

\bibitem{Forster:book}
D.~Forster.
\newblock {\em Hydrodynamic Fluctuations, Broken Symmetry and Correlation
  Functions}.
\newblock W.A. Benjamin, Inc., 1975.

\bibitem{Giuliani-Vignale:book}
G.~F. Giuliani and G.~Vignale.
\newblock {\em Quantum Theory of the Electron Liquid}.
\newblock Cambridge University Press, 2005.

\bibitem{Kubo_JPSJ57}
R.~Kubo.
\newblock Statistical-mechanical theory of irreversible processes. {I.}
  {General} theory and simple applications to magnetic and conduction problems.
\newblock {\em J. Phys. Soc. J.}, 12(6):570--586, 1957.

\bibitem{vanKampen_PN71}
N.~G. van Kampen.
\newblock The case against linear response theory.
\newblock {\em Phys. Norv.}, 5:279--284, 1971.

\bibitem{Bloch_RMP08}
I.~Bloch, J.~Dalibard, and W.~Zwerger.
\newblock Many-body physics with ultracold lattices.
\newblock {\em Rev. Mod. Phys.}, 80:885--964, 2008.

\bibitem{Enciclopedia:book}
Marcos Dantus and Peter Gross.
\newblock {\em {Ultrafast Spectroscopy in Encyclopaedia of Applied Physics}}.
\newblock Wiley, 2004.

\bibitem{Shah:book}
Jagdeep Shah.
\newblock {\em {Ultrafast Spectroscopy of Semiconductors and Semiconductor
  Nanostructures}}.
\newblock Springer, 1996.

\bibitem{Glezer:phdthesis}
Eli~Nathan Glezer.
\newblock {\em Ultrafast Electronic and Structural Dynamics in Solids}.
\newblock PhD thesis, Harvard University, Cambridge, Massachusetts, 1996.

\bibitem{Nasu:book}
Keiichiro~Nasu (ed.).
\newblock {\em {Photoinduced Phase Transitions}}.
\newblock World Scientific Publishing, 2004.

\bibitem{Shirley_PR65}
J.~H. Shirley.
\newblock Solution of schrodinger equation with a hamiltonian periodic in time.
\newblock {\em Phys. Rev.}, 138:B979, 1965.

\bibitem{Grifoni_PR98}
M.~Grifoni and P.~{H\"anggi}.
\newblock Driven quantum tunneling.
\newblock {\em Physics Reports}, 304:229--354, 1998.

\bibitem{Russomanno_PRL12}
A.~Russomanno, A.~Silva, and G.~E. Santoro.
\newblock {Periodic steady regime and interference in a periodically driven
  quantum system}.
\newblock {\em Phys. Rev. Lett}, 109:257201, 2012.

\bibitem{Landauer_PRB86}
Rolf Landauer.
\newblock Zener tunneling and dissipation in small loops.
\newblock {\em Phys. Rev. B}, 33:6497--6499, May 1986.

\bibitem{Gefen_PRL87}
Y.~Gefen and D.~J. Thouless.
\newblock Zener transitions and energy dissipation in small driven systems.
\newblock {\em Phys. Rev. Lett.}, 59:1752--1755, Oct 1987.

\bibitem{Bochner:book}
S.~Bochner and K.~Chandrasekharan.
\newblock {\em {Fourier Transforms}}.
\newblock Princeton University Press, 1949.

\bibitem{Russomanno_PRB11}
A.~Russomanno, S.~Pugnetti, V.~Brosco, and R.~Fazio.
\newblock Floquet theory of cooper pair pumping.
\newblock {\em Phys. Rev. B}, 83:214508, 2011.

\bibitem{Kollath_PRA06}
C.~Kollath, A.~Iucci, I.~McCulloch, and T.~Giamarchi.
\newblock Modulation spectroscopy with ultracold fermions in optical lattices.
\newblock {\em Phys. Rev. A}, 74:041604(R), 2006.

\bibitem{Daley_JSTAT04}
A.~J. Daley, C.~Kollath, U.~{Schollw\"ock}, and G.~Vidal.
\newblock Time-dependent density-matrix renormalization-group using adaptive
  effective hilbert spaces.
\newblock {\em JSTAT}, page P04005, 2004.

\bibitem{White_PRL04}
S.~R. White and A.~E. Feiguin.
\newblock Real-time evolution using the density matrix renormalization group.
\newblock {\em Phys. Rev. Lett.}, 93:076401, 2004.

\bibitem{Lieb_AP61}
E.~Lieb, T.~Schultz, and D.~Mattis.
\newblock Two soluble models of an antiferromagnetic chain.
\newblock {\em Annals of Physics}, 16:407--466, 1961.

\bibitem{art:Fermi-Pasta-Ulam}
E.~Fermi, J.~Pasta, and S.~Ulam.
\newblock Studies of non linear problems.
\newblock Los Alamos Report No. LA-1940, 1955.

\bibitem{Castiglione:book}
P.~Castiglione, M.~Falcioni, A.~Lesne, and A.~Vulpiani.
\newblock {\em {Chaos and coarse graining in statistical mechanics}}.
\newblock Cambridge University Press, 2008.

\bibitem{Stoeckmann:book}
Hans-J{\"u}rgen St{\"o}ckmann.
\newblock {\em {Quantum Chaos: An Introduction}}.
\newblock Cambridge University Press, 2007.

\bibitem{Sias_PRL08}
C.~Sias, H.~Lignier, Y.P. Singh, A.~Zenesini, D.~Ciampini, O.~Morsch, and
  E.~Arimondo.
\newblock Observation of photon assisted tunneling in optical lattices.
\newblock {\em Phys. Rev. Lett.}, 100:040404, 2008.

\bibitem{Lignier_PRL07}
H.~Lignier, C.~Sias, D.~Ciampini, Y.~P. Singh, A.~Zenesini, O.~Morsch, and
  E.~Arimondo.
\newblock Dynamical control of matter-wave tunneling in periodic potentials.
\newblock {\em Phys. Rev. Lett.}, 99:220403, 2007.

\bibitem{Caneva_PRB07}
Tommaso Caneva, Rosario Fazio, and Giuseppe~E. Santoro.
\newblock {Adiabatic quantum dynamics of a random Ising chain across its
  quantum critical point}.
\newblock {\em Phys. Rev. B}, 76:144427, 2007.

\end{thebibliography}
%
\end{document}